\documentclass[preprint,12pt]{elsarticle}
\usepackage{graphicx}
\usepackage{wrapfig}
\usepackage{lineno,hyperref}
\usepackage{color}
\usepackage{float}
\usepackage{amsmath,amssymb,amsthm}
\usepackage{tabularx}
\usepackage{bm}

\newcommand{\Ra}{Ra}

\newcommand{\Pra}{Pr}
\def\u{\mbox{\boldmath $u$}}
\def\f{\mbox{\boldmath $f$}}

\def\x{\mbox{\boldmath $x$}}

\def\k{\mbox{\boldmath$k$}}

\newcommand{\E}{\mathbb{E}}
\newcommand{\Prb}{\mathbb{P}}
\newcommand{\RR}{\mathbb{R}}

\newcommand{\bfU}{\mathbf{u}}

\newcommand{\DD}{\mathcal{D}}

\newcommand{\ra}{Ra}

\newtheorem{Thm}{Theorem}%[section]
\newtheorem{Lem}[Thm]{Lemma}

\newtheorem{Cor}[Thm]{Corollary}
\newtheorem{lemma}[Thm]{Lemma}
\newtheorem{corollary}[Thm]{Corollary}

\newtheorem{proposition}[Thm]{Proposition}%[section]
\newtheorem{remark}[Thm]{Remark}%[section]

\journal{arXiv.org}
\begin{document}

\begin{frontmatter}

%\shorttitle{Stability of a stochastically forced convective system}
%\shortauthor{J. F\"{o}ldes, N. Glatt-Holtz, G. Richards and J. P. Whitehead}

\title{Hydrodynamic stability in the presence of a stochastic forcing: a case study in convection}

\author[jf]{Juraj F\"{o}ldes}
\author[negh]{Nathan E. Glatt-Holtz}
\author[gr]{Geordie Richards}
\author[jpw]{Jared P. Whitehead\corref{cor1}}
\ead{whitehead@mathematics.byu.edu}

\cortext[cor1]{Corresponding author}

\address[jf]{Department of Mathematics, University of Virginia, Charlottesville, VA 22904, USA}
\address[negh]{Department of Mathematics, Tulane University, New Orleans, LA 70118, USA}
\address[gr]{Department of Mathematics \& Statistics, University of Guelph, Guelph, ON N1G 2W1, Canada}
\address[jpw]{Department of Mathematics, Brigham Young University, Provo, UT 84601, USA}

\begin{abstract}
We investigate the stability of statistically stationary conductive states for Rayleigh-B\'enard convection that arise due to a  bulk stochastic internal heating.  Our results indicate that stochastic forcing at small magnitude has little to no effect, while strong stochastic forcing has a destabilizing effect.  The methodology put forth in this article, which combines rigorous analysis with careful computation, provides an approach to hydrodynamic stability which is applicable to a variety of systems subject to a large scale stochastic forcing.
\end{abstract}

\begin{keyword}
% Stochastic Convection, Hydrodynamic Stability.
\end{keyword}

\end{frontmatter}

%\tableofcontents

%\noindent {\bf \color{red} BEGINNING OF CONTENT FROM Stability v2}

\section{Introduction}

Rayleigh B\'enard convection, the buoyancy driven motion of a fluid under the influence of a gravitational field, is ubiquitous in nature.  It is one of the driving forces in a variety of situations ranging from boiling a pot of water, to geophysical processes, to pattern formation in stellar dynamics.  Yet, despite remarkable advances in mathematical, computational, and experimental analysis,  fundamental aspects of Rayleigh-B\'enard convection remain poorly understood \citep{AhGrLo2009,ChSc2012}.

The seminal work of Lord Rayleigh \cite{Ra1916}, inspired by the experiments of B\'enard \cite{Be1900}, quantified the onset of convection in terms
of the instability of purely conductive solutions of the Boussinesq equations.  In
\cite{Ra1916} it is established that when the Rayleigh number
$\Ra$ (a dimensionless parameter proportional to the boundary heating) is less than a critical value $\Ra_c$, then the purely conductive state is globally attractive.  Not only did \cite{Ra1916} place the study of thermal convection on a firm mathematical basis, but this work also yielded a now canonical example in the study of hydrodynamic stability \cite{DrRe2004}.

Various modifications to Rayleigh's original model, including alternative boundary conditions and different sources of heat (see \cite{Go2015a} for example) have been considered in the past century. A natural further extension is to define and quantify stability for convective flows driven by stochastic forcing as many heat sources (as well as other sources of buoyancy instability) are inherently noisy.  For example, radioactive decay in the earth's mantle and thermonuclear reactions in large stars are inherently stochastic, see \cite{ScTuOl2001,KiWe1994}.

Here we investigate the hydrodynamic stability of a conductive state for a stochastic variation to the standard Rayleigh-B\'enard convection model, when an internal bulk stochastic source is present.  Our approach relies on a combination of energy stability methods, ergodic theory, and numerical computation.  It is notable that the methodology developed in this work applies to a larger class of randomly forced hydrodynamic systems which we will
address in future studies. The current investigation is also closely related to previous research by the authors on the ergodic theory and dynamical properties of stochastically
driven models for Rayleigh-B\'{e}nard convection (\cite{FoGlRiTh2015,FoGlRi2015,FoGlRiWh2016}).  Throughout the following, we provide some rigorous justification of arguments and calculations leading to our final stability result.

  While this investigation is new to the best of our knowledge, significant previous efforts have been made to incorporate
  random perturbations into models of convection.  To understand the influence of thermal fluctuations,
in \cite{GrPl1975,SwHo1977,AhCrHoSa1981,OhOrJoSeAh2004} the Boussinesq system modulated by singular (that is, active at all spatial frequencies) small noise in the bulk is investigated,
 and a reduced model is derived to describe flow statistics.  This model leads to accurate predictions of the rate of heat transfer near the onset of convection (\cite{HoSw1992}), but requires stochastic forcing stronger than the predicted thermal fluctuations.  We emphasize that these efforts were motivated by experimental observations, and the modeled noise was initiated in an effort to determine the potential influence of thermal fluctuations on carefully calibrated experiments.

 One significant difficulty in the approach initiated in \cite{SwHo1977} is that the presence of a generic stochastic source eliminates the existence of
 a traditionally defined conductive state for which the velocity field is zero,
 see \cite{GrPl1975,SwHo1977,AhCrHoSa1981,MeAhCa1991,HoSw1992,OhOrJoSeAh2004} and containing references for relevant details.  The current investigation is \emph{not} to ascertain the effect of thermal fluctuations on the onset of convection, but rather to determine the influence of stochastic effects that appear on the scale of the full system.  The goal of this investigation is to extend and further quantify a setting similar to that proposed in \cite{venturi2012} which considered the 2-dimensional Boussinesq equations with stochastic horizontal boundary conditions, and identified a substitute ``quasi-conductive regime'', for which the velocity of solutions is non-zero but small.  In \cite{venturi2012} the authors numerically showed that these quasi-conductive states maintain stability for a wide range of Rayleigh numbers containing the classical critical value $\Ra_c$.

 The current investigation will extend the results of \cite{venturi2012} to a setting where the noise is physically motivated, yet difficult if not impossible to imitate in the laboratory implying that the asymptotic and numerical investigation detailed below is necessary.  \cite{SwHo1977} and the studies that followed  addressed concerns raised by thermal fluctuations in an experimental setting. The current study is instead motivated by noise that occurs on a larger scale that is not observed experimentally, but is relevant in a practical setting as discussed below.  As such, we will not compare the results with experimental data, but infer that the instabilities suggested by the current analysis are relevant in nature and engineering applications.  Essentially the goal of this study is to evaluate the impact that a dominating, large-scale noise will have on the stability of the determined conductive state, a situation that is not realized to our knowledge in any current experimental investigation.

The numerical stability results and their rigorous justification detailed below complement the theoretical development of rigorous ergodic theorems for
stochastically forced Navier-Stokes equations, and related systems.
For example, \cite{hairer2006,hairer2008} have established that the periodic 2-dimensional Navier-Stokes equations with a bulk stochastic forcing possesses a unique ergodic invariant measure provided the stochastically forced modes satisfy a
modest geometric constraint.  These results have been extended to the stochastic Boussinesq equations by the authors
with various boundary conditions and parameter constraints (\cite{FoGlRiTh2015,FoGlRi2015,FoGlRiWh2016}).  In these settings, the unique invariant measure is almost surely globally attractive in a statistical sense.  In contrast when the stochastic forcing is ``more degenerate'', that is, when the noise is horizontally stratified then the long-time stability of statistics is much less clear when the Rayleigh number is large.  A discussion of this setting is the main goal of the present manuscript.

The starting point of the current investigation is the observation that, for certain classes of horizontally stratified volumetric forcing the system admits a dynamically inactive, conductive state $\tau(t,z)$.  This conductive solution has an explicit form and admits Gaussian statistics whose mean and covariance are readily determined.  Our primary goal in the following is to investigate the onset of convection as a bifurcation from this conductive state.

Note that our setup thus contrasts from the approach to onset taken in \cite{SwHo1977,AhCrHoSa1981,MeAhCa1991,HoSw1992} as the latter setting lacks such clearly defined conduction states.  Note furthermore that our investigations will mainly focus on fixed temperature boundary conditions with a no-slip velocity field, although comparison to the stress-free case will be made for completeness.

Our approach to this problem may be summarized as follows.  We begin by observing that the conductive state $\tau(t,z)$ (a random process dependent on the vertical spatial variable) satisfies a linear stochastic partial differential equation.  We are able to explicitly compute $\tau(t,z)$ and also the (unique) stationary distribution.  From an evolution equation for the fluctuations about $\tau(t,z)$ we derive a constrained optimization problem which provides a sufficient condition for decay of the fluctuations at an explicit random rate, denoted by $\lambda(\tau(t))$, which depends on the conductive state.  This adapts the classical energy stability method from hydrodynamic stability \citep{DrRe2004} to the stochastic setting, and analyzes the stability of the conductive state by solving a stochastic eigenvalue problem.

A crucial simplification both analytically and computationally follows because the system is stable about the conductive state provided that
\begin{equation}\label{eq:liminf_pos}
\liminf_{t\rightarrow\infty}\frac{1}{t}\int_0^t \lambda(\tau(s))ds > 0,
\end{equation}
almost surely.  Since $\tau$ is an ergodic process this expression is equivalent to integration
of $\lambda$ against the stationary law of $\tau$.
Fixing the non-dimensional stochastic heating strength $H$, and using the Dedalus computational package \citep{BuVaOiLeBrQu2017},
we identify a critical Rayleigh number  $\Ra_c$ so that \eqref{eq:liminf_pos} holds for any $\Ra\leq \Ra_c$.

We find that for small $H$ the critical Rayleigh number is comparable in value to the number obtained in \cite{Ra1916}.
However, we identify a rapid transition when the non-dimensional strength of the stochastic heating $H$ is $O(1)$ where the critical Rayleigh number $\Ra_c$ quickly decays to zero, and hence the stability of the conductive state is no longer guaranteed for any value of $\Ra$ when $H$ is sufficiently large.  Although the primary results presented here rely on computational investigation, much of the framework that underlies the variational statements can be made rigorous.  In particular, we demonstrate that the variational setup guarantees the existence of a critical Rayleigh number for which the system is stable when $\Ra< \Ra_c$.  Under certain assumptions we also justify the destabilizing effect of a strong stochastic internal heating.  The rigorous conclusions we reach are limited to a certain set of simplifying assumptions to keep the corresponding calculations tractable, but we fully anticipate that these results extend to the more interesting situation presented in the body of this paper.

The results are presented as follows: section \ref{sec:eqns_stability} introduces the equations of motion and their non-dimensionalization.  Section \ref{sec:non:lin:stab} sketches the derivation of the nonlinear stability, outlining the rigorous estimates that justify the approach and numerical results.  The proofs for these rigorous results are included in the appendices.  Section \ref{sec:numerics} discusses the numerical and algorithmic implementation of this calculation including convergence checks and criteria.  Section \ref{sec:results} contains the results including sample distributions of the critical growth factor $\lambda$.  Finally in section \ref{sec:conclusions} we draw some broad conclusions and discuss the potential extension of this method to other problems where stochasticity is present in a hydrodynamic setting.

\section{Equations of motion}\label{sec:eqns_stability}
We explore stochastic perturbations of the standard Rayleigh-B\'enard system which arises via the Boussinesq approximation:
\begin{align}\label{eq:mom_dimensional}
\frac{\partial \tilde{\u}}{\partial \tilde{t}} + \tilde{\u}\cdot\nabla\tilde{\u} + \frac{1}{\rho_0}\nabla \tilde{p} &= g\beta \k \tilde{T} + \nu \Delta \tilde{\u},\quad\quad\nabla \cdot \tilde{\u} = 0,
\end{align}
where $\tilde{\u} = (\tilde{u},\tilde{v},\tilde{w})^T$ is the three-dimensional velocity vector field, $\tilde{p}$ is the pressure, and $\tilde{T}$ is the temperature field.  In this
model the parameters are $\rho_0$ a reference density, $g$ the gravitational constant, $\beta$ the thermal expansion coefficient, and $\nu$ is the kinematic viscosity.  We are interested in a horizontally periodic box $\mathcal{D} \subset \RR^3$ of height $h$ complemented with either stress-free or no-slip boundaries for $\tilde{\u}$ on the top and bottom plates.

The temperature field $\tilde{T}$ satisfies an advection diffusion equation augmented with a stochastic forcing  in the bulk.  Stochastic forcing through the bulk is described by:
\begin{subequations}\label{eq:tem_dimensional_bulk}
\begin{align}
d\tilde{T} + \left(\tilde{\u} \cdot \nabla \tilde{T} - \kappa \Delta \tilde{T}\right)d\tilde{t} &= \gamma \sum_{k=1}^M \sigma_k dW^k,\\
 \tilde{T}(z=0) &= T_1>0,\quad \tilde{T}(z=h)=0,
\end{align}
\end{subequations}
where $\kappa$ is the thermal diffusivity, and $\gamma$ is the strength of a mean zero stochastic term that consists of $M$ independent Brownian motions $W^k$ acting on $M$ spatially orthogonal directions (in the $L^2$ norm) given by $\{\sigma_k\}$.  

Details on the mathematical setting of \eqref{eq:tem_dimensional_bulk} can be found in \cite{FoGlRiWh2016} (see \cite{ZaDa1992,KuSh2012} as well).  The limit $M\rightarrow \infty$ represents noise at all the spatial scales of the system which resembles the setting considered in \cite{SwHo1977,AhCrHoSa1981,MeAhCa1991,HoSw1992}.  Generically, we are interested in stochastic forcing on physically relevant spatial scales, that is, we will not consider forcing at scales below a given cutoff length scale.

\subsection{Non-dimensionalization, significance of parameters, and physical motivation}\label{sec:nondim}
We non-dimensionalize \eqref{eq:mom_dimensional} and \eqref{eq:tem_dimensional_bulk} by $h$ spatially, $h^2/\kappa$ temporally, and $T_1$ for the temperature.  This gives the following equivalent system (we use the same labels for the non-dimensional system, modulo the ``tilde''):
\begin{align}\label{eq:nondim_momentum}
\frac{1}{\Pra}\left(\frac{\partial \u}{\partial t} + \u\cdot\nabla\u\right) + \nabla p &= \Ra \k T + \Delta \u,\quad\quad \nabla \cdot \u= 0,
\end{align}
where the non-dimensional parameters are the Prandtl number $\Pra = \frac{\nu}{\kappa}$ capturing a kinematic property of the fluid and the Rayleigh number $\Ra = \frac{\beta g T_1 h^3}{\nu \kappa}$.  The non-dimensional temperature field for the stochastically bulk forced fluid is governed by:
\begin{subequations}\label{eq:nondim_temp_bulk}
\begin{align}
dT + \left(\u\cdot \nabla T - \Delta T\right)dt &= H \sum_{k=1}^M \sigma_k dW^k,\\
T(z=0) = 1,\quad T(z=1) = 0,
\end{align}
\end{subequations}
where the heating parameter $H = \frac{\gamma\sqrt{h}}{T_1\sqrt{\kappa}}$ is the non-dimensional ratio of the stochastic to deterministic heating.

The system evolves on the non-dimensional domain
$\mathcal{D} = [0, L]^2\times [0, 1]$ for some $L > 0$ with periodic boundary conditions in the horizontal, and either stress-free or no-slip boundaries along the top and bottom.  Analogous results are valid for the Navier-slip (Robin type condition), but they are not explored here.  Our setting allows for the unitary boundary condition on the temperature field and since the Rayleigh number is the same as in deterministic studies of convection \cite{AhGrLo2009},  we recover the system originally proposed in \cite{Ra1916} in the limit as $H \rightarrow 0$.  This is a distinctly different non-dimensionalization compared to our previous investigations of \eqref{eq:tem_dimensional_bulk}, where the relative role of the bulk stochastic heating over the deterministic boundary forcing was emphasized.  In \cite{FoGlRiWh2016} we defined two `Rayleigh parameters' $\Ra$ and $\widetilde{\Ra}$ whose product yield the Rayleigh number in this manuscript.  The reciprocal of $\widetilde{\Ra}$
from \cite{FoGlRiWh2016} yields the stochastic heating number $H$ considered here.

The Prandtl number $\Pra$ is a material property of the fluid and varies significantly  depending on the specific fluid in question.  For instance in air $\Pra \approx 0.7$, for water $\Pra\approx 7$, and analysis of the earth's mantle indicates that $\Pra \approx 10^{24}$ which is well approximated by infinity \citep{Wa2004,Wa2005,Wa2007,Wa2008b,FoGlRi2015}.  The Rayleigh number, representing the strength of the boundary driven forcing, also has a wide range in applications.  In particular the Rayleigh numbers in geophysics and astrophysics range from $10^6$ to $10^{20}$, although smaller Rayleigh numbers near the onset of convection are also of fundamental mathematical and physical interest.

\begin{table}
   \begin{minipage}{6cm}
    \begin{tabular}{cccc}
     {Physical Setting} & {$h$ (m)} & {$\kappa$ ($m^2/s$)} & {maximal $H$} \\
     Earth's mantle & $10^6$
       & $10^{-7}$ & $10^{4.5}$\\
     Earth's oceans & $10^3$    & $10^{-7}$ & $10^3$\\
     Earth's troposphere & $10^4$ & $10^{-5}$ & $10^{2.5}$\\
     Earth's convective updraft & $10^2$    & $10^{-5}$    & $10^{1.5} $\\
     convective zone in the sun & $10^8$ & $10^{-3}$ & $10^{3.5}$
    \end{tabular}
   \end{minipage}
\caption{The relevant parameters used to determine the maximal value of $H$, the stochastic heating parameter, assuming that $\frac{\gamma}{T_1}\sim 0.01$ at maximum.  The physically motivated situations here are by no means exhaustive, but they do indicate a relative maximal value of $H$ that we may motivate physically.} \label{table:H_values}
\end{table}

The heating parameter $H= \frac{\gamma\sqrt{h}}{T_1\sqrt{\kappa}}$ is the relative impact of the stochastic internal heating to the boundary driven heating, weighted appropriately by the cell height and thermal diffusivity.  The parameter $H$ also has a significant range of physically relevant values, although it is not as obvious what that range is.
The regime $H \sim 0$ occurs when the boundary forcing (characterized by $T_1$) dominates the internal stochastic heating (given by $\gamma=0$), modulated by the cell height and thermal diffusivity of the fluid.  It is difficult to compare $\gamma$ relative to $T_1$, and therefore to determine which positive, large values of $H$ are physically viable.   
However, we do expect that the stochastic effects are less significant \cite{HoSw1992}.  For our purposes we assume that the noise can be at most of the order of $\frac{\gamma}{T_1} \sim 0.01$, but this assumption is not required for our mathematical analysis.  The other two quantities $\kappa$ and $h$ are properties of the system.  Table \ref{table:H_values} displays values of parameters for several different physically relevant situations where Rayleigh-B\'enard convection is used as the first order model.  The values of $H$ are computed for $\gamma/T_1 = 0.01$, and may be adjusted if $\gamma/T_1$ changes.  The table indicates that $H$ is justifiably in the range from $0$ to $10^4$.

\subsection{The conductive state in the presence of a stochastic heat source}
\label{sec:conductive}

The conductive state for \eqref{eq:nondim_momentum} and \eqref{eq:nondim_temp_bulk} occurs when $\u = 0$.  Unlike the deterministic setting, we must retain time dependence of the temperature profile in order to modulate the stochastic forcing.  Moreover, to maintain $\u=0$, the temperature field cannot be a function of the horizontal variables, since the buoyancy term in  \eqref{eq:nondim_momentum}
needs to be absorbed into the pressure gradient.
Hence, we seek a temperature field $\tau(z,t)$ that satisfies the quasi-steady version of \eqref{eq:nondim_temp_bulk} where $\u=0$.

This indicates that $\tau(z,t)$ is a solution of
\begin{equation}\label{eq:bulk_linear}
\displaystyle{d\tau - \frac{\partial^2 \tau}{\partial z^2}dt = H\sum_{k=1}^M \sigma_k dW^k},
\end{equation}
and satisfies the non-homogeneous boundary condition $\tau(z=0)=1$ and $\tau(z=1)=0$.  To completely determine the solution to \eqref{eq:bulk_linear}, we first need to specify $\sigma_k$.  For the current investigation, we select $\sigma_k$ to be the vertically dependent eigenfunctions of the Laplace operator on the domain $\mathcal{D}$, that is, 
\begin{equation}
\sigma_k(z) = \sqrt{2} \sin(\pi k z).
\end{equation}
The function $\sigma_k$ 
is ideal for an identification of length scales in the forcing given by the vertical wave-number $k$.

The conductive state is found by separating spatial frequencies
in \eqref{eq:bulk_linear}. 
 The solution is the sum of Ornstein-Uhlenbeck processes:
\begin{align}\label{eq:conductive_time}
\tau(z,t) &= 1-z + \sqrt{2}\sum_{k=1}^M e^{-k^2\pi^2 t} \tau_k(0)\sin(\pi k z)\\ \nonumber
&+ H\sqrt{2}\sum_{k=1}^M\left[ \int_0^t e^{-k^2\pi^2 (t-s)} dW^k(s)\right]\sin(\pi k z),
\end{align}
where $\tau_k(0)$ is the coefficient of the sine series of the initial condition $\tau(z,0)$ (with subracted linear profile) corresponding to $\sigma_k$.  The stationary distribution (see \cite{ZaDa1996}) for $\tau$
in \eqref{eq:conductive_time} is given by
\begin{equation}\label{eq:bulk_stationary}
\tau^S(z) = 1-z + \sqrt{2}\sum_{k=1}^M \gamma_k \sin(\pi k z),
\end{equation}
where $\gamma_k$ are independent, normal random variables with mean $0$ and variance $\frac{H^2}{2k^2\pi^2}$, that is, $\gamma_k \sim \mathcal{N}\left(0,\frac{H^2}{2k^2\pi^2}\right)$.  We emphasize that $\tau^S$ is ergodic as a stationary solution of \eqref{eq:bulk_linear}, meaning
\begin{equation}
\label{eq:ergodic}
\lim_{t\rightarrow \infty} \frac{1}{t}\int_0^t \phi(\tau(s,\tau_0))ds = \int \phi(\tau) \mu_0 (d\tau),
\end{equation}
for any $\tau_0$ and sufficiently regular $\phi: L^2(D) \rightarrow \mathbb{R}$, where $\mu_0$ is the law of $\tau^S$ on $L^2(D)$. 
This observation is crucial in the analysis that follows.

\section{Nonlinear stability of the conductive state}\label{sec:non:lin:stab}

In this section we formulate  a sufficient condition for stability of the stochastic conductive state $\tau$ identified in \eqref{eq:conductive_time}
for  systems  with  stochastic bulk forcing.  In Subsection \ref{sec:nonlinstab:1}, we show that, a sufficient condition for stability is that a stochastic growth factor $\lambda$ (defined below in \eqref{eq:lambda_variational_min}) is positive when integrated against the stationary distribution $\tau^{S}$ of the conductive state, that is, $\tau$ is almost surely stable provided that $\E(\lambda(\tau^S))>0$.

The growth factor $\lambda=\lambda_{Pr}(Ra,\tau)$ depends on $Pr$ and $Ra$, as well as the conductive state $\tau$, and the conductive state $\tau$ depends on the stochastic forcing parameter $H$.  We first provide rigorous foundations (under certain simplifying assumptions) for the variational approach, demonstrating the existence of a critical $Ra_c$, and proving that sufficiently strong stochastic forcing is destabilizing.  In Sections \ref{sec:numerics}--\ref{sec:results}, we numerically approximate a critical Rayleigh number $Ra_{c}$ at fixed $\Pra$ such that for $Ra<Ra_c$ the sufficient condition for stability, $\E(\lambda_{Pr}(Ra,\tau^S))>0$, is satisfied.  The numerical results also quantify how $Ra_{c}$ varies with the forcing parameters and demonstrates how the distribution of $\lambda_{Pr}$ is decidedly non-Gaussian even in the marginal case  $\E(\lambda_{Pr}) \approx 0$.  %Subsection \ref{sec:nonlinstab:2} is devoted to the statement of rigorous results related to the growth factor $\lambda$ which provide a partial theoretical foundation for our numerical analysis strategy.

\subsection{Energy stability and a stochastic variational problem}\label{sec:nonlinstab:1}
Since the conductive state is time dependent, we will not consider linear stability but will focus entirely on nonlinear stability via the energy method (see \cite{DrRe2004} and \cite{Go2015a,Go2015b} for example).  Specifically, 
we decompose the temperature field as $T(x,y,z,t) = \tau(z,t) + \theta(x,y,z,t)$ so that \eqref{eq:nondim_momentum} and \eqref{eq:nondim_temp_bulk} become
\begin{align}\label{eq:mom_perturb}
\frac{1}{\Pra}\left(\frac{\partial \u}{\partial t} + \u\cdot\nabla\u\right) + \nabla \hat{p} &= \Ra \k \theta + \Delta \u,\quad\quad \nabla \cdot \u = 0,\\ \label{eq:theta_perturb}
\frac{\partial\theta}{\partial t} + \u \cdot\nabla \theta + w\frac{\partial \tau}{\partial z} &= \Delta\theta,
\end{align}
where the pressure term $\hat{p}$ has been modified to absorb the buoyancy term from the conductive profile $\tau$.
Note that this system is stochastic only through the presence of $\tau$, and in particular the perturbed system obeys the rules of ordinary calculus.
We compute the evolution of the energy ($L^2$ norm of $\theta$ and $\u$) as
\begin{multline}%\label{eq:energy_evolution}
\frac{1}{2}\frac{d}{dt}\left(\|\theta\|^2 + \frac{1}{\Pra\Ra}\|\u\|^2\right) = -\mathcal{Q}(\u,\theta,\tau),\\ \label{eq:define_Q}
\mbox{where} \quad\mathcal{Q}(\u,\theta,\tau) = \|\nabla\theta\|^2 + \frac{1}{\Ra}\|\nabla \u\|^2 + \int_\mathcal{D} w\theta\left(\frac{\partial \tau}{\partial z} -1\right)d\x
\end{multline}
 and we define the $L^2$ norm as $\|f\|^2 = \|f\|_2^2 = \int_\mathcal{D} |f|^2d\x$.

For fixed $\tau$, let $\mathcal{Q}$ be a quadratic form in $\u$ and $\theta$, and following the energy stability method \citep{DrRe2004} we consider
\begin{equation}\label{eq:lambda_variational_min}
\displaystyle{\lambda(\tau) = \min_{\u,\theta}\frac{\mathcal{Q}(\u,\theta,\tau)}{\|\theta\|^2+(\Pra \Ra)^{-1}\|\u\|^2}},
\end{equation}
 which is a random quantity depending on the parameters $\Pra,~\Ra,$ and $H$.  For a more precise formulation see details in Subsection \ref{sec:nonlinstab:2} below.  The Euler-Lagrange equations for the minimization problem \eqref{eq:lambda_variational_min}
  with Lagrange multipliers $\lambda$ that enforces normalization, and $q(\x)$ that guarantees incompressibility, are:
 \begin{align}\label{eq:v-eqn}
 \frac{\lambda}{\Pra\Ra} \u &= -\frac{1}{\Ra}\Delta \u + \nabla q + \k\frac{1}{2}\left(\frac{d\tau}{dz} - 1\right) \theta,\\
 \nabla \cdot \u &= 0,\\ \label{eq:th-eqn}
 \lambda \theta &= -\Delta \theta + \frac{1}{2}\left(\frac{d\tau}{dz} - 1\right) w.
 \end{align}
It is important to notice that by testing 
\eqref{eq:v-eqn} and \eqref{eq:th-eqn} by $\u$ and $\theta$ respectively, and using that $\u$ is 
incompressible, we obtain 
\begin{eqnarray*}
   \lambda\Big(\|\theta\|^2+ \frac{1}{\Pra \Ra}\|\u\|^2\Big) = \mathcal{Q}(\u,\theta,\tau) \,.
\end{eqnarray*}
In particular, the minimum $\lambda(\tau)$ from 
\eqref{eq:lambda_variational_min} is the smallest eigenvalue (Lagrange multiplier) of \eqref{eq:v-eqn}--\eqref{eq:th-eqn}. 
 
 From \eqref{eq:define_Q} and \eqref{eq:lambda_variational_min} we conclude that $\tau$ is a unique and exponentially asymptotically stable state of \eqref{eq:nondim_momentum} coupled with \eqref{eq:nondim_temp_bulk} provided that
\begin{equation}\label{eq:tauS_gtr0}
\liminf_{t\rightarrow \infty} \frac{1}{t}\int_0^t \lambda(\tau(s))ds > 0.
\end{equation}
  Invoking geometric ergodicity \citep{ZaDa1996}, with rigorous justification discussed in more detail below in Subsection \ref{sec:nonlinstab:2}, we have that
\begin{equation}\label{eq:tauS_ergodicity}
\lim_{t\rightarrow \infty}\frac{1}{t}\int_0^t \lambda(\tau(s))ds = \mathbb{E} \lambda(\tau^S),
\end{equation}
almost surely, independent of the initial condition, where $\tau^S$ is the stationary distribution of the conductive state and $\mathbb{E}$ denotes the statistical mean.  We conclude that a sufficient condition for almost sure exponential stability of the conductive state $\tau$ is $\mathbb{E} \lambda(\tau^S) >0$.

\subsection{Rigorous analysis of the variational formulation} \label{sec:nonlinstab:2}
We will consider a range of physically plausible values for the stochastic forcing parameter $H$, and numerically approximate a `critical' Rayleigh number $\Ra=\Ra_c$ so that $\mathbb{E}\lambda(\tau^S) \approx 0$.  In this subsection, we state rigorous results in support of 
\begin{enumerate}
    \item[(i)] $\mathbb{E}\lambda(\tau^S)>0$ provided that $Ra < Ra_c$.
    \item[(ii)] a strategy for estimating $\Ra_c$ by a Monte Carlo algorithm.
\end{enumerate}
The proofs of results stated in this subsection can be found in \ref{app:rigbounds}.

There are several instances in this subsection where simplifying assumptions are imposed to make the resulting estimates more tractable.  We formulate these assumptions when appropriate, but emphasize that we expect that most of the additional assumptions are \emph{not} necessary for the stated results, just necessary for obtaining tractable proofs.

We start with definitions and notation.
\begin{enumerate}
\item We seek to analyze \eqref{eq:lambda_variational_min} for $(\u,\theta) \in \mathcal{H}$, where
\begin{equation}\label{eq:define_space}
\mathcal{H} = \left\{(\u,\theta): \nabla \cdot \u = 0,\quad \u \in H_0^1(\DD)^3,\quad \theta \in H_0^1(\DD)\right\}
\end{equation}
and $H_0^1(\DD)$ is the usual Sobolev space of functions with square integrable gradient and zero boundary conditions. 
\item To establish rigorous estimates on \eqref{eq:tauS_ergodicity} we first prove bounds on \eqref{eq:lambda_variational_min} for a fixed $\tau(z)$.  In the following, we will use $\eta(z)$ to refer to a fixed realization of the conductive state, and only use $\tau(z)$ when we are referring to the coupled Ornstein-Uhlenbeck processes defined in Section \ref{sec:conductive}. Specifically, $\tau(z)$ refers to a random variable while $\eta(z)$ is one specific deterministic realization of $\tau(z)$.
\item Let $\mathcal{Y} $ be the set of all linear combinations of $\sin(k\pi z)$, that is, 
\begin{equation}\label{eq:define_Y}
\mathcal{Y} = \mbox{span}\{\sin(k\pi z),\mbox{where }k = 1,\ldots M \},
\end{equation}
where $M$ is the number of modes forced in \eqref{eq:bulk_linear}.
\item Since \eqref{eq:lambda_variational_min} is invariant under the scaling $(\u,\theta)\rightarrow (a\u,a\theta)$, then \eqref{eq:lambda_variational_min} is equivalent to
\begin{align}
\label{eq:var_new}
\lambda_{\Pra}(\Ra,\eta) &= \inf_{(\u,\theta)\in \mathcal{M}} Q(\u,\theta,\eta),\\
 \mbox{where } \label{eq:variational_constraint}
 \quad \mathcal{M} &= \left\{(\u,\theta) \in \mathcal{H}: \|\theta\|_2^2 + \frac{1}{\Pra\Ra}\|\u\|_2^2 = 1\right\},
\end{align}
where $\mathcal{Q}$ is given by \eqref{eq:define_Q}.
\end{enumerate}

We begin by establishing the existence of a minimizer and regularity of $\lambda$ with respect to $\eta$. In the following, we denote $W^{1,\infty}([0,1])$ the space of Lipchitz functions, or equivalently the space of functions with almost everywhere bounded derivatives. 

\begin{Thm} \label{thm:3-1}
Let $\eta:[0, 1] \to \RR$ be a deterministic function. 
\begin{enumerate}
\item If $\eta \in W^{1,\infty}([0,1])$, then there exists a minimizer $(\u,\theta) \in \mathcal{H}$ for the variational problem \eqref{eq:var_new}.
\item For every fixed $\Ra$ and $\Pra$, the map $\lambda_{\Pra}(\Ra,\cdot): W^{1,\infty}([0, 1]) \rightarrow \mathbb{R}$ as defined by \eqref{eq:var_new} is globally Lipschitz.
\end{enumerate}
\end{Thm}

A detailed proof of Theorem \ref{thm:3-1} is found in \ref{sec:rigor_prereq}.  With Theorem \ref{thm:3-1} (and related lemmata established during its proof) at our disposal, we are prepared to establish the sufficiency of the stability condition identified in Subsection \ref{sec:nonlinstab:1}, and to prove that the notion of a critical Rayleigh number extends to the stochastically bulk-forced setting.

\begin{Thm}\label{thm:3-2}
The zero solution of \eqref{eq:mom_perturb}-\eqref{eq:theta_perturb} is exponentially asymptotically stable almost surely if
\begin{equation}
\E \lambda_{\Pra}(\Ra,\tau^S) > 0.
\end{equation}
Furthermore,
\begin{equation}
\E \lambda_{\Pra}(\overline{\Ra},\tau^S) < \E \lambda_{\Pra}(\Ra,\tau^S)
\end{equation}
whenever $\overline{\Ra}>\Ra$.  Thus there exists at most one $\Ra_c$ such that $\E\lambda_{\Pra}(\Ra,\tau^S) > 0$ for $\Ra < \Ra_c$ and $\E\lambda_{\Pra}(\Ra,\tau^S) < 0$ for $\Ra > \Ra_c$.
\end{Thm}

The proof of Theorem \ref{thm:3-2} is presented in \ref{sec_rigor_exist}.  From Theorem \ref{thm:3-2} we conclude a critical Rayleigh number $\Ra_c$ can be obtained by finding the root of $\E\lambda_{\Pra}(\Ra_c,\tau^S) = 0$, but we require numerical methods to approximate this root and quantify its dependence on the forcing parameters.  The remainder of the theorems presented in this section are devoted to analysis of the growth factor $\lambda$ in order to provide insight into how $\Ra_c$ can be estimated in various parameter regimes.

In particular, our next objective is to investigate the functional dependence of the growth factor $\lambda = \lambda_{\Pra}(\Ra,\tau^S)$ on the forcing coefficients $\gamma = (\gamma_1,\ldots,\gamma_M)$ defined in \eqref{eq:bulk_stationary}.  To simplify notation, we define
\begin{equation}\label{eq:define_zeta}
\zeta(\gamma) = \sum_{k=1}^M k\pi \cdot \gamma_k \cos(k \pi z),
\end{equation}
where each component $\gamma_k$ is independently distributed, with 
\begin{equation}\label{eq:gamma_dist}
\gamma_k \sim \mathcal{N}\left(0,\frac{H^2}{2k^2\pi^2}\right).
\end{equation}
Then by \eqref{eq:bulk_stationary} we have $\partial_z \tau^S = \zeta(\gamma)-1$, allowing us to re-interpret $\lambda_{\Pra}(\Ra,\tau^S)= \lambda_{Pr}(Ra,\zeta(\gamma))$\footnote{Note that we consider $\gamma \in \RR^M$ as a new variable.} using \eqref{eq:define_Q} and \eqref{eq:var_new}.  We can further simplify notation in subsequent statements, for fixed $Ra$ and $Pr$, by denoting $\lambda(\gamma):= \lambda_{Pr}(Ra,\zeta(\gamma))$ for input deterministic forcing coefficients $\gamma\in\mathbb{R}^M$.
We establish the following results.

\begin{Thm}
\label{thm:3-3}
The function $\gamma \mapsto \lambda(\gamma):\RR^n \to \RR$ is continuous, concave, and has one sided directional derivatives, in particular for any $\nu \in \mathbb{R}^M$ one has
\begin{align*}
\partial_\nu^+\lambda(\gamma) &:= \lim_{h\rightarrow 0^+} \frac{\lambda(\gamma+h\nu)-\lambda(\gamma)}{h}
= \inf_{(\u^*,\theta^*)\in\mathcal{Z}} \int_\mathcal{D} w^*\theta^* \zeta(\nu),\\
\partial_{\nu}^-\lambda(\gamma) &:= \lim_{h\rightarrow 0^-}\frac{\lambda(\gamma+h\nu)-\lambda(\gamma)}{h}
= \sup_{(\u^*,\theta^*)\in\mathcal{Z}}\int_{\mathcal{D}}w^*\theta^*\zeta(\nu),
\end{align*}
where $\mathcal{Z}$ is the set of all global minimizers of the variational problem \eqref{eq:var_new} with functional $Q(\u,\theta,\zeta(\gamma))$.
In addition, for any $\nu \in \mathbb{R}^M$,
\begin{equation}
\label{eq:der_bound}
\left|\partial_\nu^{\pm}\lambda(\gamma)\right| \leq \frac{\pi \sqrt{\Ra\Pr}}{2} \sum_{k=1}^M k|\nu_k|.
\end{equation}
\end{Thm}

One consequence of Theorem \ref{thm:3-3} is that  $\gamma \mapsto \lambda(\gamma)=\lambda_{\Pra}(\Ra,\zeta(\gamma))$ is  continuous, and therefore the expected value $\E\lambda_{\Pra}(\Ra,\tau^S) = \E\lambda_{\Pr}(\Ra,\zeta(\gamma))$ can be obtained by integrating $\lambda(\gamma)$ against the law of $\gamma$.  The proof of Theorem \ref{thm:3-3} is presented in \ref{sec:rigor_concave}.

Next, we seek bounds on $\gamma \mapsto \lambda(\gamma)$ in order to estimate expectation in \eqref{eq:tauS_ergodicity}, which consequently yields estimates on the critical Rayleigh number $\Ra_c$.  First we estimate $\lambda(0)$, which can be computed explicitly, as this is the growth factor for the standard deterministically forced Rayleigh-B\'enard problem \cite{Ra1916}. To avoid implicit formulas and to have an explicit set of eigenfunctions of the Stokes operator, we consider stress free boundary conditions on the horizontal plates:
\begin{eqnarray}\label{bc:str-free}
    \partial_z u(x, y, \pm 1, t) = 
    \partial_z v(x, y, \pm 1, t) = 
    w(x, y, \pm 1, t) = 0 \,.
\end{eqnarray}
As usual, we assume periodic conditions in the horizontal direction of the domain $\mathcal{D} = [0, L]^2\times[0, 1]$. 
Then, we derive estimates on the derivative $\partial_\nu^+\lambda(0)$ which yield a tangent approximation at 0, and due to concavity of $\gamma \mapsto \lambda(\gamma)$, the approximation is 
an upper bound on $\lambda(\gamma)$.
The proof of the following theorem is given in  \ref{sec:rigor_find_gammac}. To illustrate the ideas, we restrict to a special choice of Prandtl number $Pr$, however we expect that the same conclusions are valid in general. 

\begin{Thm}\label{thm:3-4}  For simplicity of computations we assume that $Pr=1$, $L<2\sqrt{2}$, and stress free boundary conditions \eqref{bc:str-free}.
\begin{enumerate}
\item If $\ra \leq   \frac{4\pi^4(4 + L^2)}{L^4}$, then  for any $\nu \in \RR^{M}$,
\begin{equation}
\label{eq:lamb_est_1}
\lambda(0) = \pi^2 \quad \textrm{and } \quad \partial_{\nu}^+ \lambda(0) = 0\,.
\end{equation}
\item If $ \frac{4\pi^4(4 + L^2)}{L^4} \leq \ra \leq \frac{\pi^4 (4 + L^2)^3}{4L^4}$, then
\begin{align}
\label{eq:lam_est2}
\begin{split}
\lambda(0) &= \pi^2\left(1 + \frac{4}{L^2} \right) - \frac{2\sqrt{\ra}}{\sqrt{4 + L^2}} \,, \\ &\textrm{and } \quad
\partial_{\nu}^+ \lambda(0) =  -\frac{\sqrt{\ra}}{\sqrt{4 + L^2}}\frac{\nu_2}{2\pi}.
\end{split}
\end{align}
\item If $ \ra > \frac{\pi^4 (4 + L^2)^3}{4L^4}
$, then  $\lambda (\ra, \gamma) \leq 0$ for each $\gamma \in \mathbb{R}^{M}$.
\end{enumerate}
\end{Thm}
Observe that the derivative $\partial_{\nu}^+ \lambda(0)$ is discontinuous at $\ra = \frac{\pi^4 (4 + L^2)^3}{4L^4}$ which is due to the fact that $L$ is finite, and therefore the spectrum of the operator is discrete. 

Let us illustrate heuristic consequences that emerge from Theorem \ref{thm:3-4}.  First, for $Pr=1$ and $L<2\sqrt{2}$, it is clear from part 3 of Theorem \ref{thm:3-4} combined with Theorem \ref{thm:3-2} that the critical Rayleigh number satisfies $\ra_c < \frac{\pi^4 (4 + L^2)^3}{4L^4}$.  Then, from parts 1 and 2 of Theorem \ref{thm:3-4} we have $\lambda(0) \geq 0$ and $\partial_\nu^+ \lambda(0) \leq 0$, and therefore by the concavity of $\gamma \mapsto \lambda (\gamma)$
we find that for any fixed $\gamma\in\mathbb{R}^M$, the function $\varepsilon \mapsto \lambda(\varepsilon\gamma)$ is non-increasing for $\varepsilon>0$.

\begin{figure}[!htb]
\begin{center}
\scalebox{.6}{\includegraphics[trim = 1mm 1mm 1mm 1mm, clip]{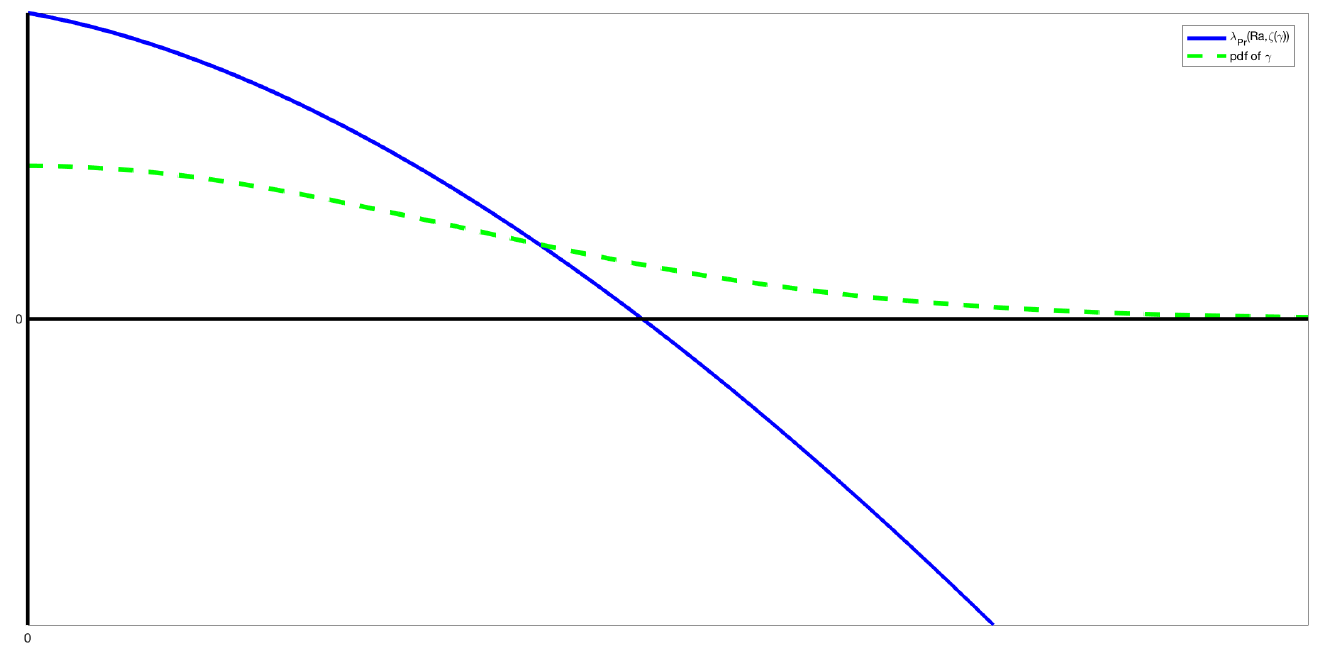}}
\end{center}
  \caption{A cartoon depiction of the growth factor dependence on $\gamma$.  On the horizontal axis is $\gamma \in \mathbb{R}$, although the same concepts apply for $\gamma\in \mathbb{R}^M$.  For evenly forced modes \eqref{eq:lambda_variational_min} is symmetric in $\gamma$ so we only illustrate $\gamma\geq 0$.  The solid (blue online) curve is an illustration of $\lambda_{\Pr}(\Ra,\zeta(\gamma))$ as a function of $\gamma$.  In particular note that $\lambda_{\Pr}(\Ra,\zeta(\gamma))$ is concave everywhere, and as $\gamma\rightarrow\infty$ asymptotes to a linear function.  The dashed (green online) curve represents the normal distribution from which $\gamma$ is drawn.  $\E[\lambda_{\Pr}(\Ra,\zeta)]$ is then computed by integrating the product of these two functions.}
\label{fig:lambda_vs_pdf}
\end{figure}

For fixed $\gamma \in \RR^M$, Figure \ref{fig:lambda_vs_pdf} depicts $\varepsilon \mapsto \lambda(\varepsilon\gamma)$ in the solid (blue) plot. 
The dashed line depicts the probability distribution of $\gamma$ given in \eqref{eq:gamma_dist} projected to the one-dimensional subspace spanned by $\gamma$.  If $M=1$, that is, $\gamma$ is one dimensional, the critical Rayleigh number $\Ra_c$ is the value of $\Ra$ such that the integral of the product of the curves in Figure \ref{fig:lambda_vs_pdf} is equal to zero. An analogous principle holds in higher dimensions, that is, when $M>1$.  

Let us investigate the dependence of $\ra_c$ on the strength of the forcing $H$. 
If $H$ increases, then the variance of the Gaussian distribution increases, meaning that the dashed curve in Figure \ref{fig:lambda_vs_pdf}  widens (recall \eqref{eq:gamma_dist}). Because the solid line (independent of $H$) is decreasing, $\E\lambda_{\Pra}(\Ra,\zeta(\gamma))$ decreases. 
  On the other hand, decreasing $\Ra$ will raise the solid curve in Figure \ref{fig:lambda_vs_pdf}, increasing $\E\lambda_{\Pra}(\Ra,\zeta(\gamma))$ (recall Theorem \ref{thm:3-2}), and therefore 
to keep $\E\lambda_{\Pra}(\Ra_c,\zeta(\gamma))=0$, 
we have to decrease the variance of the distribution, which means that 
  $\Ra_c=\Ra_c(H)$ \textit{is a decreasing function of $H$}.

\begin{figure}
\begin{center}
\scalebox{.6}{\includegraphics[trim = 1mm 1mm 1mm 1mm, clip]{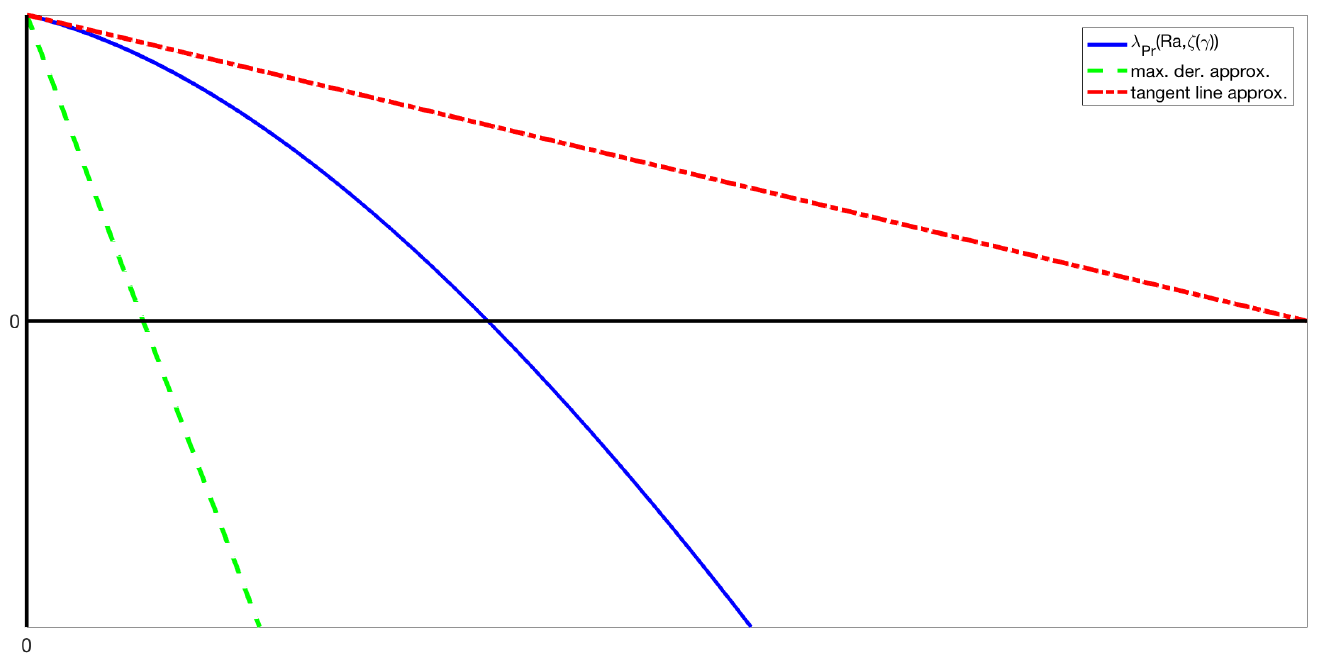}}
\end{center}
  \caption{A cartoon depiction of the dependence of the growth factor on $\gamma$.  The axes are the same as in Figure \ref{fig:lambda_vs_pdf} and the solid (blue online) curve is the same.  The upper dash-dot (red online) line represents the tangent line approximation at $\gamma=0$ and provides an upper bound on the growth factor, and an upper bound estimate on the intersection point $\gamma^*$ wherein $\lambda_{\Pr}(\Ra,\zeta(\gamma^*)) = 0$.  The dashed lower (green online) line represents the lower bound obtained by considering the linearization with the maximally negative derivative and provides a lower bound estimate on $\gamma^*$.  Note that the exact value of $\gamma^*$ is not estimated well in this illustration, but at asymptotically small or large values of $H$ such approximations are still quite useful.}
\label{fig:lambda_1}
\end{figure}

For $ \frac{4\pi^4(4 + L^2)}{L^4} \leq \ra \leq \frac{\pi^4 (4 + L^2)^3}{4L^4}$, we combine parts 2 and 3 of Theorem \ref{thm:3-4} to obtain a decreasing tangent line approximation from above for the function $\varepsilon \mapsto \lambda(\varepsilon\gamma)$, for any $\gamma\in\mathbb{R}^M$.  Specifically,  a lower bound is obtained from the linear approximation having maximal negative derivative (in any direction) given by \eqref{eq:der_bound}.  These two linear approximations illustrated in Figure \ref{fig:lambda_1} provide us with an upper and lower bound on  the growth factor $\lambda$.

\begin{figure}
\begin{center}
\scalebox{.6}{\includegraphics[trim = 1mm 1mm 1mm 1mm, clip]{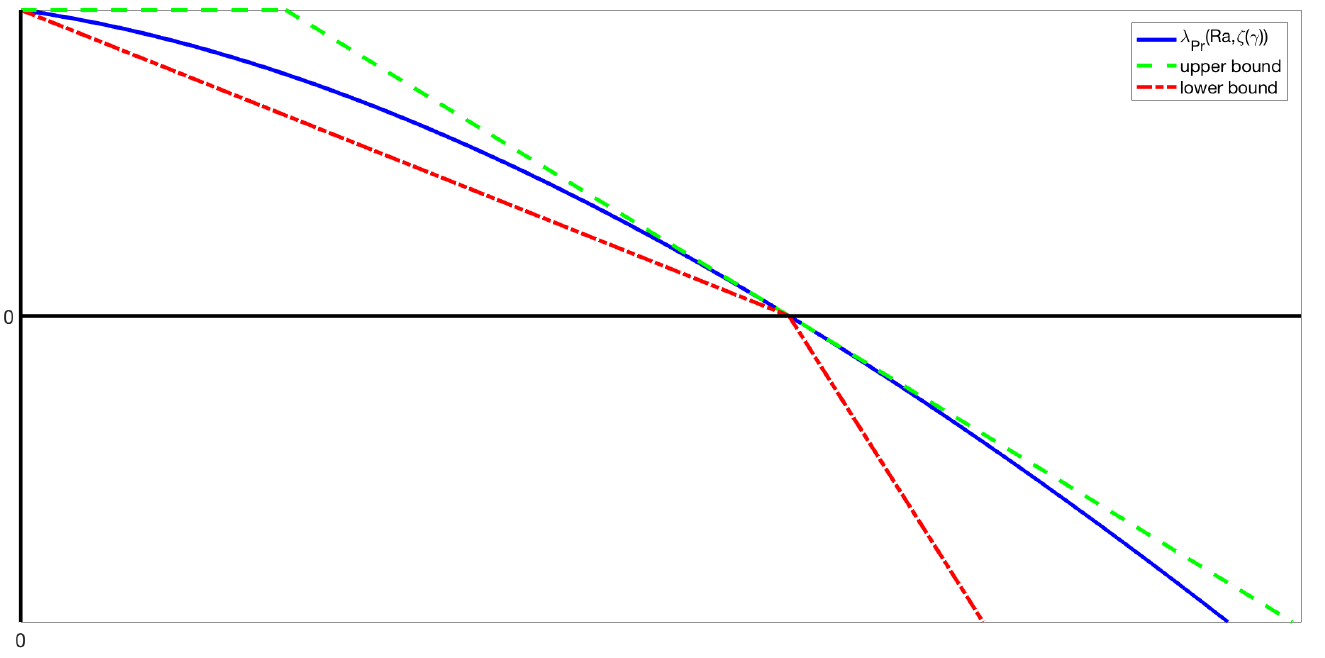}}
\end{center}
  \caption{A cartoon depiction of the dependence of the growth factor on $\gamma$. The axes are the same as in Figure \ref{fig:lambda_vs_pdf} and the solid (blue online) curve is the same.  The dashed upper (green online) lines represent upper bounds on $\lambda_{\Pr}(\Ra,\zeta(\gamma))$ as described in the text, and the dash-dot (red online) lines represent the lower bounds.}
\label{fig:lambda_2}
\end{figure}

\ref{sec:rigor_H_dep} finally contains 
refined upper and lower bounds on the 
growth factor $\lambda$ an their integral against the Gaussian distribution of $\gamma$.  Figure \ref{fig:lambda_2} illustrates how  we obtained these refined bounds.  The upper bound is first obtained by applying a tangent approximation at the point $\gamma=\gamma^*$ (the point at which the growth factor is zero) cut off by the maximal value of the growth factor which is obtained at $\gamma=0$. 
Recall that this is indeed an upper bound, since $\lambda$ is non-increasing and concave. 

The lower bound is also obtained as a piece-wise linear function.  First, a linear interpolation between the points at $\gamma=0$ and $\gamma=\gamma^*$ is used for the region where the growth factor is positive.  Then, a linear approximation using the maximally negative derivative originating at $\gamma=\gamma^*$ is used as a lower bound for negative values of $\lambda$.  Overall, we use upper and lower bounds to estimate the dependence of the critical Rayleigh number $\Ra_c$ on the stochastic forcing parameter $H$.  

First, the lower bound (depicted in Figure \ref{fig:lambda_2}) is used to show that for small values of the internal heating $H>0$ the system is more stable (the critical Rayleigh number is larger with the addition of noise) than the corresponding deterministic forcing.  This is stated under certain simplifying assumptions (only meant to provide an accessible proof, we do not anticipate that these restrictions are necessary) in the following theorem.

\begin{Thm}\label{thm:lower_bounds}
Assume that $\mathcal{N}=\{2\}$ with the corresponding coefficient $\gamma := \gamma_2$, $\Pr=1, L \leq 2\sqrt{2}$, and boundary conditions \eqref{bc:str-free}. Denote (cf. Theorem \ref{thm:3-4})
\begin{equation}
    \Ra_L := \frac{4\pi^4(4+L^2)}{L^4} \qquad 
    \Ra_U := \frac{\pi^4 (4 + L^2)^3}{4L^4} 
\end{equation}
Then for any $\overline{\Ra} \in [\Ra_L, \Ra_U]$
there exists unique $\overline{\gamma} > 0$  such that $\lambda_{\Pr=1}(\overline{\Ra},\zeta(\overline{\gamma})) = 0$ (recall that $\gamma \in \RR$ is a scalar and $\zeta : \RR \to \RR$). 
However, $\E_{\gamma}[\lambda_{\Pr=1}(\overline{\Ra},\zeta(\gamma))] > 0$, provided $\gamma = N_H \sim\mathcal{N}\left(0,\frac{\gamma_H^2}{8\pi^2}\right)$.  In other words the deterministic system is marginally stable at $(\overline{\Ra}, \overline{\gamma})$, but the stochastically forced system is almost surely stable for the corresponding internal heating.
\end{Thm}

Finally we state the final Theorem here that partially justifies the computational calculations provided below, i.e. finite sampling of the desired distribution will provide a viable approximation of the actual expected value.

\begin{Thm}\label{thm:gaussian_tails}
Assume that $\Pr=1$ in addition to the other assumptions incorporated in the previous Theorem.  It follows that
\begin{align}
&\frac{1}{2}\left(2\lambda_L - \Ra^{1/2} - \Ra^{1/2}\sqrt{2}\pi \sum_{k\in\mathcal{N}}k|\gamma_k|\right) \leq \lambda_{\Pr=1}(\Ra,\zeta(\gamma))\\
&\leq \frac{1}{2}\min_k \left(\pi^2 k^2 - \frac{\sqrt{2}}{4}\Ra^{1/2} \left[\sqrt{2}\pi k|\gamma_k| + 2\right]\right).
\end{align}
Hence, taking the expectation of this inequality, and because $\gamma_k$ are Gaussian (and hence so is $|\gamma_k|$) we see that $\lambda_{\Pr=1}(\Ra,\zeta(\gamma))$ must have Gaussian tails for $\lambda << 0$ ($\lambda$ is bounded from above).
\end{Thm}

The presence of Gaussian tails justifies the computational evidence discussed below (Monte Carlo simulations of phenomena with Gaussian tails are well justified as the low probability events can safely be neglected computationally).

\section{Algorithmic description, numerical comparisons, and rigorous convergence}\label{sec:numerics}

The computation of the marginally stable parameters, i.e. those parameters for which $\E\lambda_{Pr} (\ra,\tau^S) = 0$
is performed as follows.  This is a root-finding problem where the Prandtl number $\Pra$ is fixed and the stochastic heating $H$ and total number of forced modes $M$ are fixed parameters.  The Rayleigh number $\Ra$ is the independent variable.  We approximate $\mathbb{E}\lambda({\tau^S})$ from a sample mean, and seek $\Ra_c$ so that $\widehat{\lambda}(\Ra_c) = 0$, where
\begin{equation}
\label{eq:sample_mean}
\widehat{\lambda}(\Ra) = \frac{1}{\mathcal{N}}\sum_{k=1}^{\mathcal{N}} \lambda (\tau^S_k),
\end{equation}
 and each $\tau^S_k$ is drawn independently from the stationary distribution of the conductive state.  The number of samples
$\mathcal{N}$ is a parameter that we select.  As $\lim \mathcal{N}\rightarrow \infty$ is the desired situation, we will take $\mathcal{N}$ as large as practical computational considerations will allow.  Due to numerical considerations, we also must seek $\Ra_c$ such that
$|\widehat{\lambda}(\Ra_c)| \leq \epsilon$, where $\epsilon$ is also selected via numerical considerations.

For each realization $\tau^S_k$ of the stationary distribution, we consider the Euler-Lagrange equations for the minimization  of $\lambda(\tau)$ and identify $\lambda(\tau^S_k)$ as the solution of a one-dimensional eigenvalue problem which is solved numerically via the Dedalus software package \citep{BuVaOiLeBrQu2017}.
Starting with the Euler-Lagrange equations
\eqref{eq:v-eqn}--\eqref{eq:th-eqn}
for $\lambda(\tau)$, we twice take the 
$\textrm{curl}$ of \eqref{eq:v-eqn} and using the fact that 
$\textrm{curl}\textrm{curl} = \nabla \textrm{div} - \Delta$ we have 
\begin{align}
 \frac{\lambda}{\Pra\Ra} \Delta w &= -\frac{1}{\Ra}\Delta^2 w  + \frac{1}{2}\left(\frac{d\tau}{dz} - 1\right) \Delta_H \theta,\\ 
 \lambda \theta &= -\Delta \theta + \frac{1}{2}\left(\frac{d\tau}{dz} - 1\right) w \,,
 \end{align}
where $w$ is the third (vertical) component of the velocity and $\Delta_H = \partial^2_x + \partial^2_y$ 
is the horizontal Laplacian. 
Next, we decompose $w$ and $\theta$ into horizontal Fourier series 
\begin{eqnarray*}
    w(x, y, z) = \sum_{k \in \mathbb{Z}_L^2} \hat{w}_k(z)e^{ik\cdot(x, y)} \qquad
\theta(x, y, z) = \sum_{k \in \mathbb{Z}_L^2} \,, \hat{\theta}_k(z)e^{ik\cdot(x, y)}    
\end{eqnarray*}
where $\mathbb{Z}_L^2 = \{2\pi k/L: k \in \mathbb{Z}^2\}$, to obtain:
\begin{align}\label{eq:four-w}
-\frac{\lambda}{\Pra\Ra}\left(\partial^2_{zz}-|k|^2\right) \hat{w}_k &= \frac{1}{\Ra}\left(\partial^2_{zz} - |k|^2\right)^2\hat{w}_k + \frac{1}{2}\left(\partial_z \tau - 1\right) |k|^2 \hat{\theta}_k,\\\label{eq:four-th}
\lambda \hat{\theta}_k &= -\left(\partial^2_{zz}-|k|^2\right)\hat{\theta}_k + \frac{1}{2}\left(\partial_z\tau - 1\right)\hat{w}_k \,.
\end{align}
Since our operators are self-adjoint, $\lambda$ and all coefficients
$\hat{w}_k$ and $\hat{\theta}_k$ are real. 
 The system also satisfies for each $k \in \mathbb{Z}^2_L$ the boundary conditions:
\begin{equation}
\hat{w}_k(0) = \hat{w}_k(1)=\partial_{z}\hat{w}_k(0)=\partial_{z}\hat{w}_k(1)=0,\quad\quad \hat{\theta}_k(0)=\hat{\theta}_k(1)=0,
\end{equation}
for no-slip boundaries, where we used the identity $u = v = 0$ on $\partial D$, which implies $u_x = v_y = 0$ and by the incompressibility condition, $w_z = 0$ on $\partial D$.  
Similarly, if $\u$ satisfies the stress-free boundary condtions, then $w = 0$, and $\partial_z u = \partial_z v = 0$ on $\partial D$, and consequently 
$\partial_{xz} u = \partial_{yz} v = 0$. Differentiating the divergence free condition $\nabla \cdot \u = 0$ with respect to $z$, implies 
$\partial_{zz}w = - \partial_{xz} u - \partial_{yz} v = 0$ on $\partial D$, and consequently
\begin{equation}
\hat{w}_k(0) = \hat{w}_k(1)=\partial^2_{zz}\hat{w}_k(0)=\partial^2_{zz}\hat{w}_k(1)=0,\quad\quad \hat{\theta}_k(0)=\hat{\theta}_k(1)=0,
\end{equation}
for stress-free boundaries.

\begin{figure}
\begin{center}
\includegraphics[width=\textwidth]{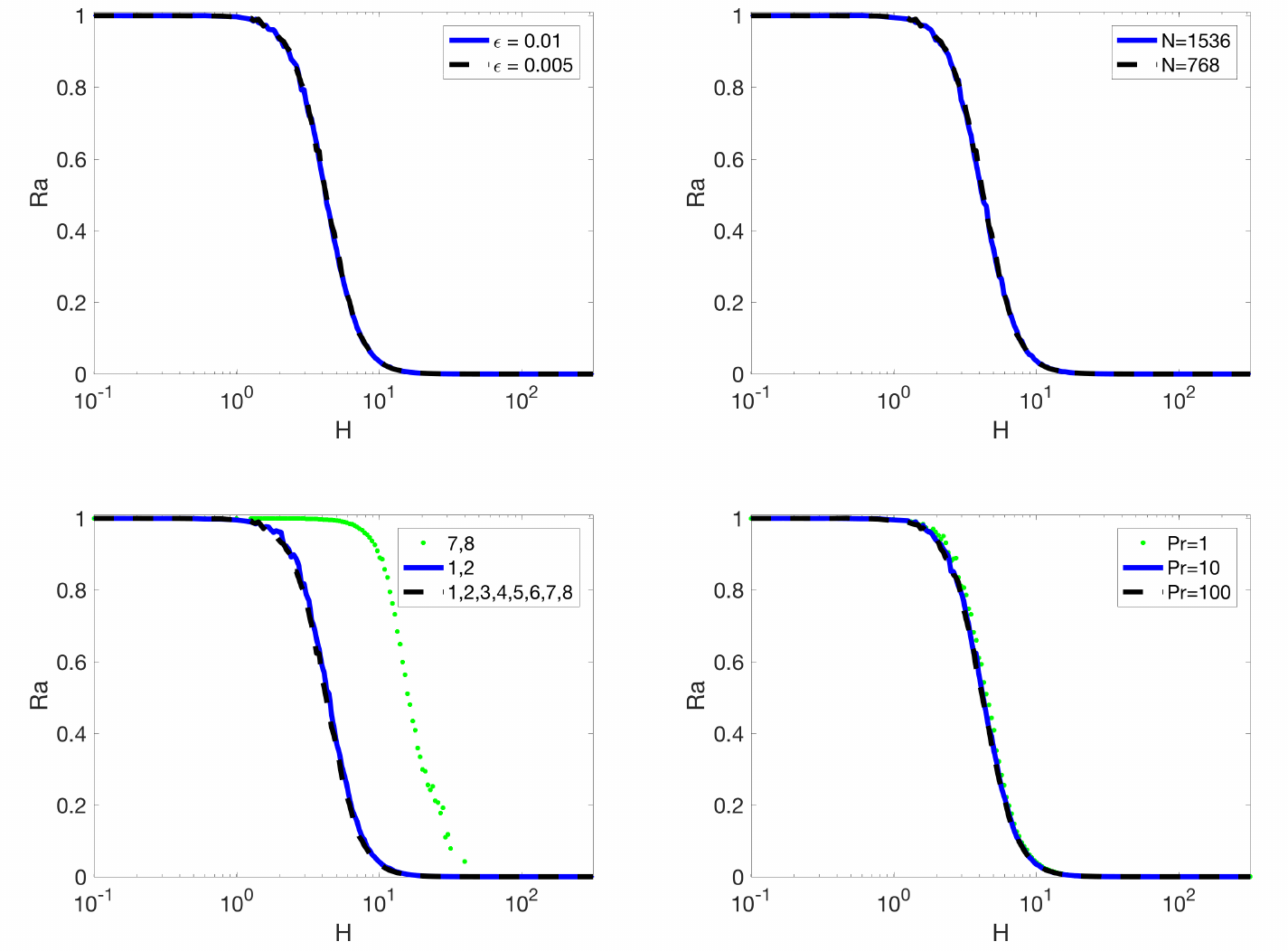}
\end{center}
  \caption{Comparisons of the numerical results for the bulk stochastic heat source under a variety of different numerical and physical configurations.  For each plot, the vertical axis is the critical Rayleigh number normalized by its deterministic value.  The horizontal axis represents different values of the non-dimensional heating parameter $H$.  All the reported cases in this Figure are computed for no-slip boundaries.  The default values as indicated in the text are $\epsilon=0.005,~\mathcal{N}=768,~M=1,2,3,4,5,6,7,8,~$ and $\Pra=10$.  In each of these comparisons, only one of these parameters is varied at a time, with all others held fixed at their default values.}
\label{fig:numerical_comparison}
\end{figure}

The numerical implementation uses a modification of the bisection root finding method on
$\widehat{\lambda}$ to identify the approximate $\Ra_c$, for each specified value of $\Pra$ and the pair $(H,M)$ by taking previously computed values for nearby parameter values as an initial guess.  This leads to several parameters in the algorithm that must be selected (numerical comparisons are displayed in Figure \ref{fig:numerical_comparison}) for a variety of different parameters.

\subsubsection*{Numerical details}
\begin{enumerate}
\item \emph{Absolute error}.  Using $\epsilon=0.005$ as the absolute tolerance allows for a determination of $\Ra_c$ at $H=0$ that is accurate up to 8 significant digits.
We select $\epsilon = 0.005$ for the results reported here.  The upper left plot in Figure \ref{fig:numerical_comparison} shows the difference between the calculated critical Rayleigh number between $\epsilon=0.005$ and $\epsilon=0.01$ with the other parameters set to their default values as described below.  The maximal difference between the critical Rayleigh number for $H\leq 10$ is less than $5\%$, jumping to a maximum of $12\%$ for the higher less physical values of $H$.  All other calculations reported here are for $\epsilon=0.005$.
\item \emph{Sample size}.  We choose $\mathcal{N}=768$ Monte Carlo generated samples to compute $\widehat{\lambda}$.  As a control, similar calculations were performed for $\mathcal{N}=1536$, but as shown in the upper right plot of Figure \ref{fig:numerical_comparison} the differences are minimal particularly for $H \lesssim 10$.  The maximal difference in this comparison is less than $9\%$.  All other parameters are the default values as explained in this section.
\item \emph{Forced modes}.  We choose to force the first 8 modes as the default setup.  The lower left plot in Figure \ref{fig:numerical_comparison} demonstrates the effects of varying the modes that are forced with all other parameters set as default values.  The differences here are far more significant, with the primary conclusion being that the choice of the lowest forced mode is the most influential on the stability calculation.  When only the $7$ and $8$ vertical modes are forced at sufficiently large values of $H$, the viscous effects are not able to control the asymptotically strong, small scale oscillations causing the Monte Carlo sampling to fail to converge.  This explains the jagged nature of the plot in this figure for $H > 10$.  In addition, the root finding algorithm was unable to converge for $\epsilon=0.005$ and $\mathcal{N}=768$ for $H$ larger than that shown in this plot.  We expect that increasing the sample size by an order of magnitude would eliminate this issue, but would also be far to computationally prohibitive to be useful.
\item \emph{Choice of the Prandtl number}.  The lower right plot in Figure \ref{fig:numerical_comparison} shows differences for variations on $\Pra$ for the default parameters.  There is certainly some dependence on $\Pra$ in this case, particularly when comparing $\Pra=1$ and $\Pra=100$, but these changes do not qualitatively alter the primary conclusions, and hence we choose $\Pra=10$ as the default value.
\item \emph{Level of discretization}.  We chose to discretize the eigenvalue problem with $N_z=64$ vertical Chebyshev modes.  Results are identical up to 6 significant digits for $N_z=128$ as long as the highest forced mode was less than 8, i.e. $M \leq 8$.  If the highest forced mode was chosen higher than $8$ then the vertical discretization would need to be increased accordingly.
\item \emph{Boundary conditions}.  As the no-slip boundary condition is the most physically relevant to modern experiments, we focus on this boundary condition, but we have also performed comparisons with the stress-free boundary condition and found very little dependence on the velocity boundary condition.  Although the actual values of $\Ra_c$ are significantly different for the different boundary conditions, $\Ra_c(H)/\Ra_c(0)$ and within the tolerance we have admitted, the exact same transition.
\end{enumerate}

In summary, the default parameters for the simulations are: $\epsilon = 0.005,~\mathcal{N} = 768,~\Pra=10,~N_z = 64$, and the first 8 modes of the bulk are forced with no-slip boundaries.

\section{Results}\label{sec:results}
The primary takeaway from the numerical results is that weak stochastic forcing has a stabilizing effect, essentially retaining the same critical Rayleigh number for small to moderate values of $H$ as when $H=0$.  There is then a rapid transition (dependent on the number of modes forced for the bulk heat source) as $H$ is increased whereon the system is strongly destabilized as $\Ra_c\rightarrow 0$.

\begin{wrapfigure}{l}{0.6\textwidth}
\begin{center}
\includegraphics[width=0.6\textwidth]{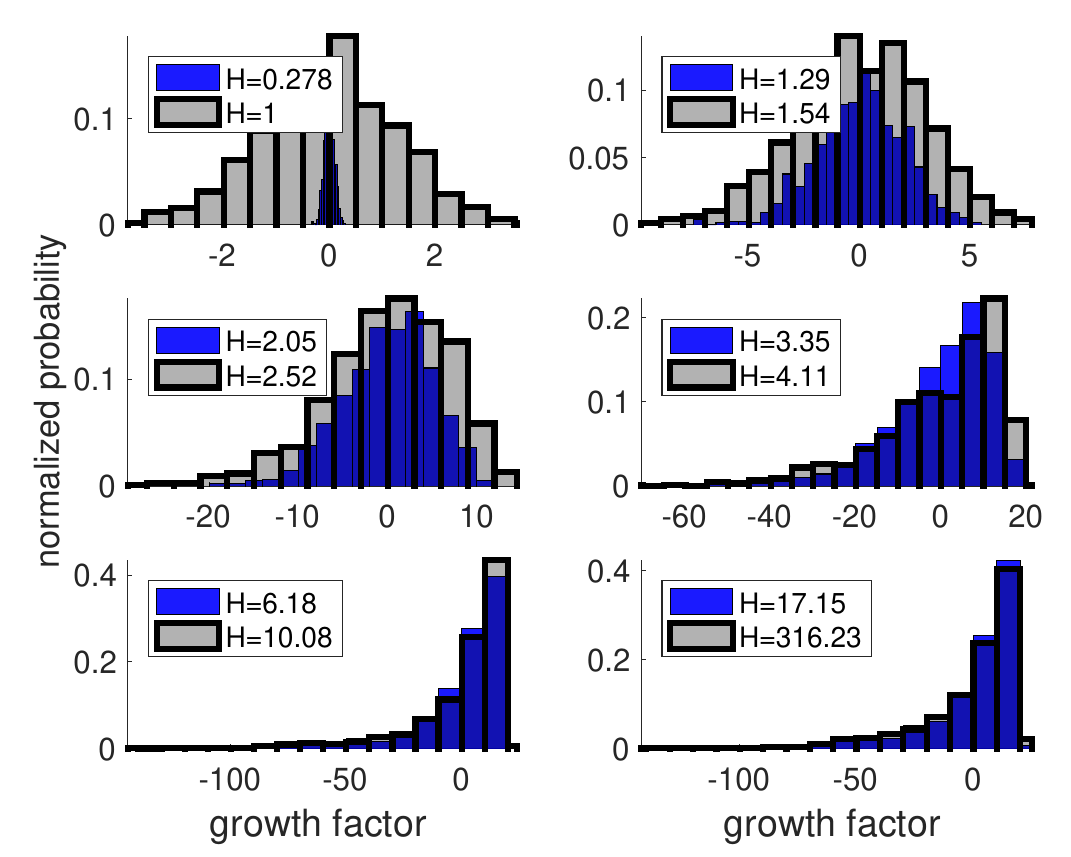}
\end{center}
  \caption{Normalized histograms of the growth factor $\lambda$ for the 768 samples that identified the critical Rayleigh number for a variety of values of $H$.  All reported results used $\epsilon=0.005,~\mathcal{N}=768,~\Pra=10,~N_z=64$ and the first 8 vertical modes of the bulk were forced.  Note the difference in the horizontal scale between each plot.}
\label{fig:bulk_histograms}
\end{wrapfigure}

To fully investigate the transition from the conductive to the convective regime, we consider the distribution of the growth factors $\lambda$ from the 768 samples for each value of $H$ at the transitional value of $\Ra = \Ra_c$ as demonstrated in Figure \ref{fig:bulk_histograms}.  We have chosen these particular values of $H$ as they represent different qualitative settings for the onset of instability.  The variance of $\lambda$ is clearly and unsurprisingly an increasing function of $H$, but numerically it appears that the maximal value of $\lambda$ is bounded from above (for rigorous proof in a special case see Lemma \ref{lem:gauss_tail}), causing the distribution of $\lambda$ to skew strongly to the left as $H$ increases.  By definition, the mean of these histograms must remain zero for all values of $H$ so that as the distribution skews negatively then $\Ra_c$ must decrease toward zero rather suddenly in $H$.

\begin{wrapfigure}{l}{0.6\textwidth}
\begin{center}
\includegraphics[width=0.6\textwidth]{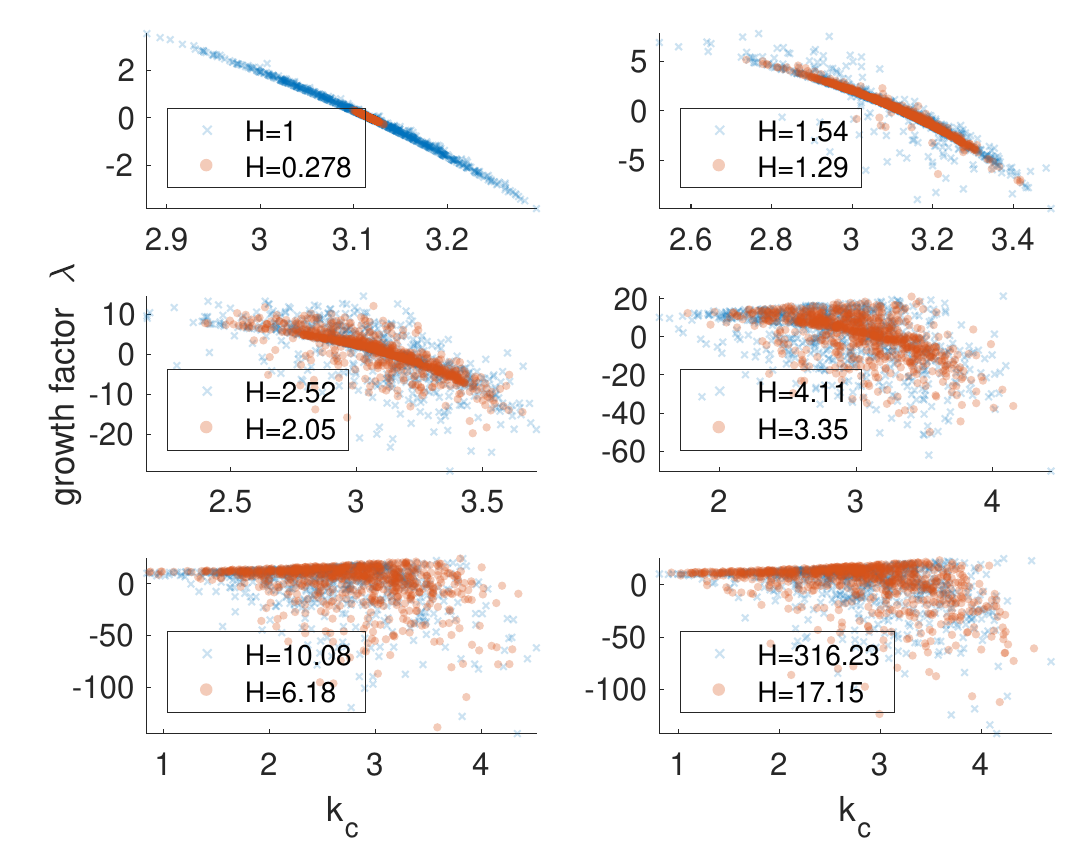}
\end{center}
  \caption{Scatter plots of the critical horizontal wave number $k_c$ against the corresponding growth factor $\lambda$ at the critical state, i.e. the mean of $\lambda$ is 0.  All reported results used $\epsilon=0.005,~\mathcal{N}=768,~\Pra=10,~N_z=64$ and the first 8 vertical modes of the bulk were forced.  Note the difference in scale between each plot.}
\label{fig:bulk_scatter}
\end{wrapfigure}

A more detailed picture regarding the appearance of these instabilities comes from considering Figure \ref{fig:bulk_scatter}.  Figure \ref{fig:bulk_scatter} shows the dependence of the growth factor $\lambda$ on the critical horizontal wavenumber $k_c$ which indicates the spatial scale at which the instability will arise.  For smaller values of $H$ the results are physically intuitive.  The more unstable modes (negative values of $\lambda$) arise at the larger wave-numbers, i.e. smaller scales.  This indicates that when the noise is weak, there is a small-scale instability which naturally may arise as a result of the inherent randomness of the stochastic forcing.  As the stochastic forcing increases in strength, this relationship breaks down to some degree insomuch that for $H \geq 6.18$ the unstable modes occur for $k_c \gtrsim 1.5$ with the most unstable modes typically being at higher wave-numbers.

The upper bound on $\lambda$ is also clearly seen in Fig. \ref{fig:bulk_scatter} for $H \geq 3.35$, and interestingly it appears that there is also an upper bound on $k_c$ so that even for the extreme case of $H=316.23$, the maximal $k_c$ either stable or unstable is less than $5$.  These bounds on both $\lambda$ and $k_c$ in the presence of a stochastic forcing with increasing variance (with respect to $H$) leads to an increasingly varied scattering of both $\lambda$ and $k_c$ below the upper bounds, i.e. an extremely non-symmetric distribution.  Even so, the largest scales indicated by $k_c < 1.5$ remain stable.

Another interesting feature that emerges from Fig. \ref{fig:bulk_scatter} is the nearly linear relationship between $\lambda$ and $k_c$ for smaller values of $H$.  Although there is significant variation from this behavior for $H \geq 1$, the same banded structure still appears in each of the plots shown in Fig. \ref{fig:bulk_scatter}.  This feature is indicative of the instability that arises in the deterministic setting when $H=0$ (near the known critical value of $k_c\approx 3.1$), with the stochastic noise yielding stability for larger scales and instability for smaller ones.  As $H$ increases this behavior still holds, but an additional trend appears wherein stable modes appear for nearly all wave-numbers below the cut-off of $k_c=5$, and as the variance of the noise increases dramatically, the scatter about these two trends increases as well.

\section{Conclusions and stability of generic stochastically driven hydrodynamic systems}\label{sec:conclusions}
We have investigated the nonlinear stability of a convective system with additive stochastic white noise on the first 8 vertical modes of the system.  When the stochastic heating is weak, it has a stabilizing effect, then transitions to strongly destabilizing as the strength of the noise increases.  This is an effect of the distribution of the growth factor $\lambda$ and its dependence on the strength of the stochastic heating.

The results discussed above have demonstrated the need to better quantify the role that stochasticity plays in physically relevant fluid systems.  If the internal heat source were modeled as a deterministic bulk forcing, then the stability of the resultant conductive state would be very different, particularly for the physically relevant setting of $H = o(1)$.  This implies that at least in this idealized convective setting, if there is an inherently noisy source, we will miss some of the fundamental physics by modeling the system in a purely deterministic fashion.  We are not claiming here that the precise nature of the noise we have chosen is the `best' way to model noisy convection, but we do insist that accounting for noise in such physical systems is necessary to achieve physically realistic results.

Further considerations of the onset of convection in such stochastic settings are natural extensions of the current work.  For instance, is there an analogue of the finite amplitude equations in this context, and if so, is their derivation and application the same or similar?  Do coherent structures such as the roll-states present in low Rayleigh number deterministic convection exist in the stochastic setting, and if so are they defined only in a mean sense?  If such structures exist, can similar statements be made regarding their stability, or does the stochastic nature of the problem preclude the utility of such investigations?  Further analysis and computation is required to answer these questions, and ascertain the influence that noise can play in fully developed turbulent convection.

Finally, we note that the methodology developed in this article applies not only to Rayleigh-B\'enard convection under these constraints, but is applicable to any hydrodynamic system driven by a stochastic forcing where a basic (time dependent) state still exists.  In particular, we can extend this approach to shear flow, particularly where there is noise present in the induced shear boundary condition, or Rayleigh-B\'enard convection when there is noise introduced through the temperature boundary condition.

%\begin{appendix}

\appendix

\section{Proofs of Rigorous Bounds on the Growth Factor}\label{app:rigbounds}

\subsection{A priori estimates and necessary prerequisites}\label{sec:rigor_prereq}
We begin with an estimate that proves existence of a solution to the minimization problem and indicates that there are indeed cases such that $\lambda_{\Pra}(\Ra,\eta) < 0$.

\begin{lemma}\label{lem:existence}
For a fixed $\eta \in W^{1,\infty}([0,1])$ there exists a minimizer $(\u,\theta) \in \mathcal{H} \setminus \{0\}$ of \eqref{eq:var_new} where $\mathcal{H}$ is given by \eqref{eq:define_space}.  In addition, if $\eta \in \mathcal{Y}$ (see \eqref{eq:define_Y} for the definition of $\mathcal{Y}$) then there exists a constant $C(\Ra,\Pra)>0$ so that if $\|\eta'\|_2 \geq C(\Ra,\Pra)$, then $\lambda_{\Pra}(\Ra,\eta)<0$, where $\eta'$ is the first derivative of $\eta$.
\end{lemma}

\begin{proof}
We note that for $(\u,\theta) \in \mathcal{M}$ (see \eqref{eq:variational_constraint}), there is a constant $C=C(\Pra,\Ra)$ such that
\begin{eqnarray*}
\int_\mathcal{D} w\theta (\partial_z\eta - 1)d\x
&\leq& \left(\|\eta\|_{W^{1,\infty}}+1\right)\|\u\|_2\|\theta\|_2\\
&\leq& C (\|\eta\|_{W^{1,\infty}}+1)\left(\|\theta\|_2^2 + \frac{1}{\Pra\Ra}\|\u\|_2^2\right)\\
&=&  C(\|\eta\|_{W^{1,\infty}} + 1),
\end{eqnarray*}
implying that $Q(\u,\theta,\eta)$ is bounded from below.  We note that $Q(\u,\theta,\eta)$ contains a part which is equivalent to the $H^1$ norm of $\u$
and $\theta$ and a part which is compact, and therefore $Q(\u,\theta,\eta)$
is weakly lower semicontinuous in $(\u, \theta)$ for any fixed $\eta$. In addition, $\mathcal{M}$ is weakly closed (the embedding of $\mathcal{H}$ into $(L^2)^4$ is compact) in the topology of $H^1$.  Then from the direct methods in the calculus of variations, there exists a minimizer of \eqref{eq:lambda_variational_min}.

Fix any compact set $K\subset \mathcal{D}$ and let $\chi_K$ be a smooth function compactly supported in $\mathcal{D}$ so that $\chi_K = 1$ on $K$. The set $\mathcal{Y}_K = \{\phi \chi_K : \phi \in \mathcal{Y}\}$ is a finite dimensional space of trigonometric polynomials restricted to $K$, then all norms on $\mathcal{Y}_K$ are equivalent. In particular,

\begin{equation*}
\|\eta_K\|_{H^3} \leq c_0\|\eta_K'\|_2,
\end{equation*}
for any $\eta_k \in \mathcal{Y}_K$. 
Choose $\tilde{\u} = \left(x\eta_K'',0,-\eta_K'\right)^T$ and $\tilde{\theta} = |\eta_K'|$ as test functions in \eqref{eq:lambda_variational_min}.  Then $(\tilde{\u},\tilde{\theta})$ is compactly supported in $\mathcal{D}$, $\tilde{\u}$ is divergence free, and therefore admissible. In particular, 
\eqref{eq:lambda_variational_min} and the equivalence of norms  show that
\begin{eqnarray*}
\lambda_{\Pra}(\Ra,\eta) &\leq& C \frac{\|\eta_K''\|_2^2 + \frac{1}{\Ra}\left(\|\eta_K'''\|_2^2 + 2\|\eta_K''\|_2^2\right)-\|\eta_K'\|_{L^3}^3 + \|\eta_K'\|_2^2}{\|\eta_K''\|_2^2 + \|\eta_K'\|_2^2}\\
&\leq& C(1-c\|\eta_K'\|_2),
\end{eqnarray*}
where the constants $C$ and $c$ depend only on $\Pra,~\Ra,$ and $\mathcal{D}$.  The second assertion in the statement of the lemma follows immediately if $\|\eta_K'\|_2$ is sufficiently large.
\end{proof}

\begin{remark}
Uniqueness of the specified minimizer is not guaranteed, and in fact is not true in general.  We can show that there is at most a one-parameter family of minimizers for each $\eta$.  As this does not reflect on the results that follow the details are omitted here.
\end{remark}

In all of the following we will also need the following lemma which asserts the smoothness of the map $\eta \mapsto \lambda_{\Pra}(\Ra,\eta)$.

\begin{lemma}\label{lem:lambda_Lip}
For every fixed $\Ra$ and $\Pra$, the map $\lambda_{\Pra}(\Ra,\cdot): W^{1,\infty}([0, 1]) \rightarrow \mathbb{R}$ as defined in \eqref{eq:lambda_variational_min} is globally Lipschitz.
\end{lemma}

\begin{proof}
Fix $\overline{\eta}\in W^{1,\infty}(0,1)$ and let $(\overline{\u},\overline{\theta})\in \mathcal{M}$ be the minimizer of \eqref{eq:lambda_variational_min} with $\tau$ replaced with $\overline{\eta}$ (for the existence of such minimizer see Lemma \ref{lem:existence}).  For any $\eta\in W^{1,\infty}(0,1)$ we find that
\begin{align*}
\lambda_{\Pra}(\Ra,\overline{\eta}) &= \mathcal{Q}(\overline{\u},\overline{\theta},\overline{\eta})\\
&\geq \mathcal{Q}(\overline{\u},\overline{\theta},\eta) - \|\overline{\eta}-\eta\|_{W^{1,\infty}}\|\overline{\u}\|_2\|\overline{\theta}\|_2\\
&\geq \mathcal{Q}(\overline{\u},\overline{\theta},\eta) - C\|\overline{\eta}-\eta\|_{W^{1,\infty}}\\
&\geq \lambda_{\Pra}(\Ra,\eta) - C\|\overline{\eta}-\eta\|_{W^{1,\infty}},
\end{align*}
where $C$ is a constant depending on $\Ra>0$.
%(see the final lines in the proof of Lemma \ref{lem:existence} above).  
Then, 
\begin{equation}
\lambda_{\Pra}(\Ra,\eta) - \lambda_{\Pra}(\Ra,{\overline{\eta}}) \leq C\|\overline{\eta}-\eta\|_{W^{1,\infty}},
\end{equation}
which upon interchanging $\eta$ and $\overline{\eta}$ gives the statement of the lemma.
\end{proof}

We frequently make use of the following result which indicates properties obeyed by the minimizer of \eqref{eq:lambda_variational_min}.

\begin{lemma}\label{lem:cross_term_neg}
Fix $\eta \in W^{1,\infty}([0,1])$ and let 
$(\overline{\u},\overline{\theta})$ be a minimizer of \eqref{eq:lambda_variational_min} as in 
Lemma \ref{lem:existence} with $\tau$ replaced by $\eta$. Then
\begin{equation}\label{eq:cross_term_neg}
\int_\mathcal{D}\overline{w}\overline{\theta}(\partial_z\eta - 1) d\x \leq 0.
\end{equation}
Moreover, if $\lambda_{\Pra}(\Ra,\eta) \leq 0$, then
\begin{equation}\label{eq:cross_term_neg:2}
\int_\mathcal{D}\overline{w}\overline{\theta}(\partial_z\eta - 1) d\x \leq - c_0 \min\{\Pr, 1\},
\end{equation}
where $c_0$ depends only on $\mathcal{D}$. 
In particular if $\eta \in \mathcal{Y}$ and $\|\eta'\|_2 \geq C$ as described in Lemma \ref{lem:existence}, then \eqref{eq:cross_term_neg:2} holds.
\end{lemma}

\begin{proof}
As $(\overline{\u},\overline{\theta})$ is a minimizer of $\lambda_{\Pra}(\Ra,\eta)$, then
\begin{equation}
\frac{\mathcal{Q}(\overline{\u},\overline{\theta},\eta)}{\|\overline{\theta}\|_2^2 + \frac{1}{\Pra\Ra}\|\overline{\u}\|_2^2} \leq \frac{\mathcal{Q}(\overline{\u},-\overline{\theta},\eta)}{\|\overline{\theta}\|_2^2+\frac{1}{\Pra \Ra}\|\overline{\u}\|_2^2},
\end{equation}
or equivalently
\begin{equation}
\int_\mathcal{D}\overline{w}\overline{\theta}(\partial_z\eta - 1)d\x \leq 0.
\end{equation}
If $\lambda_{\Pra}(\Ra,\eta) \leq 0$, then by the Poincar\' e inequality there exists $c$ depending only on $\mathcal{D}$ such that
\begin{equation}\label{eq:sting}
\begin{aligned}
  \int_\mathcal{D}\overline{w}\overline{\theta}(\partial_z\eta - 1) d\x &\leq 
  -\|\nabla \overline{\theta}\|^2 - \frac{1}{\Ra} \|\nabla \overline{\u}\|^2 
  \leq - c \Big(\|\overline{\theta}\|^2 + \frac{1}{\Ra} \|\overline{\u}\|^2 \Big) \\
  &\leq -c\min\{\Pr, 1\} \Big(\|\overline{\theta}\|^2 + \frac{1}{\Ra\Pr} \|\overline{\u}\|^2 \Big)
  \\
  &\leq -c\min\{\Pr, 1\}  \,,
\end{aligned}
\end{equation}
as desired. 
\end{proof}

\subsection{The existence of a critical Rayleigh number}\label{sec_rigor_exist}
The goal of the present subsection is to prove the following proposition:
\begin{proposition}\label{prop:stability}
The zero solution of \eqref{eq:mom_perturb}-\eqref{eq:theta_perturb} is exponentially asymptotically stable almost surely if
\begin{equation}
\E \lambda_{\Pra}(\Ra,\tau^S) > 0.
\end{equation}
Furthermore, there is $c = c(\Pr) > 0$ such that for any $\Ra^*>\Ra$ we have
\begin{equation}\label{eq:stexp}
\E \lambda_{\Pra}(\Ra^*,\tau^S) \leq \E \lambda_{\Pra}(\Ra,\tau^S) - c(\sqrt{\Ra^*} - \sqrt{\Ra})
\end{equation}
and therefore $\Ra \mapsto \E \lambda_{\Pra}(\overline{\Ra},\tau^S)$ is strictly decreasing. Also, for any  $\Ra^*, \Ra > 0$, we obtain
\begin{equation}\label{eq:idnye}
|\E\lambda_{\Pra}(\Ra,\tau^S) - \E\lambda_{\Pra}(\Ra^*,\tau^S)| \leq C \max\{\Pr, 1\} |\sqrt{\Ra} - \sqrt{\Ra^*}|\,,   
\end{equation}
and in particular $\Ra \mapsto \E \lambda_{\Pra}(\Ra,\tau^S)$ is continuous.

Thus there exists at most one $\Ra_c$ such that $\E\lambda_{\Pra}(\Ra,\tau^S) > 0$ for $\Ra < \Ra_c$ and $\E\lambda_{\Pra}(\Ra,\tau^S) < 0$ for $\Ra > \Ra_c$.
\end{proposition}

First, however, 
for each $\eta \in W^{1, \infty}([0, 1])$
we need to establish the monotonicity and continuity of $\Ra \mapsto \lambda_{\Pr}(\Ra,\eta)$ and hence the existence of a critical $\Ra_c$ for any fixed $\eta\in W^{1,\infty}([0,1])$.

\begin{lemma}\label{lem:lambda_monotonic_Ra}
For a given $\eta \in W^{1,\infty}([0,1])$ and $\Pra>0$, the map $\Ra \mapsto \lambda_{\Pra}(\Ra, \eta):[0,\infty)\rightarrow \mathbb{R}$ is decreasing and 
\begin{equation}\label{eq:swcnt}
    |\lambda_{\Pra}(\Ra,\eta) - \lambda_{\Pra}(\Ra^*,\eta)| \leq \|\eta\|_{W^{1, \infty}([0, 1])} \max\{\Pr, 1\} |\sqrt{\Ra} - \sqrt{\Ra^*}|.
\end{equation}
Moreover if $\eta \in \mathcal{Y}$ (see \eqref{eq:define_Y} for the definition), 
with $\|\eta'\|_2\geq C$, where $C = C(\Ra^*,\Pra)$ given in Lemma \ref{lem:existence}, then 
\begin{equation}\label{eq:sosmp}
  \lambda_{\Pra}(\Ra,\eta) \geq   \lambda_{\Pra}(\Ra^*,\eta) + c \min\{1, \Pr\} (\sqrt{\Ra} - \sqrt{\Ra^*}) \,,
\end{equation}
where $c = c(\mathcal{D})$. In particular, if $\|\eta'\|_2$ is large, then $\Ra \mapsto \lambda_{\Pra}(\Ra,\eta)$ is strictly decreasing. 
\end{lemma}

\begin{proof}
Fix any $\eta\in W^{1,\infty}([0,1])$ and $\Ra^*\geq \Ra \geq 0$.  Let $(\sqrt{\Ra}\overline{\u},\overline{\theta})$ be the minimizer of \eqref{eq:lambda_variational_min} as in Lemma \ref{lem:existence}. 
Then using Lemma \ref{lem:cross_term_neg}, $\Ra \leq \Ra^*$, 
\begin{align*}
\lambda_{\Pra}(\Ra,\eta) &= \frac{\|\nabla\overline{\theta}\|_2^2 + \|\nabla \overline{\u}\|_2^2 + \sqrt{\Ra}\int_\mathcal{D}\overline{w}\overline{\theta}(\partial_z \eta - 1) d\x}{\|\overline{\theta}\|_2^2 + \frac{1}{\Pra}\|\overline{\u}\|_2^2}\\
&= \frac{\|\nabla\overline{\theta}\|_2^2 + \|\nabla\overline{\u}\|_2^2 + \sqrt{\Ra^*}\int_\mathcal{D}\overline{w}\overline{\theta}(\partial_z \eta-1)d\x}{\|\overline{\theta}\|_2^2  + \frac{1}{\Pra}\|\overline{\u}\|_2^2}
\\
&\qquad
+ \frac{(\sqrt{\Ra} - \sqrt{\Ra^*})\int_\mathcal{D}\overline{w}\overline{\theta}(\partial_z \eta-1)d\x}{\|\overline{\theta}\|_2^2 + \frac{1}{\Pra}\|\overline{\u}\|_2^2}
\\
&\geq \inf_{(\u,\theta)\in \mathcal{H}} \frac{\mathcal{Q}_{\Ra^*}(\u,\theta,\eta)}{\|\theta\|_2^2+\frac{1}{\Pra\Ra^*}\|\u\|_2^2}\\
&= \lambda_{\Pra}(\Ra^*,\eta) \,,
\end{align*}
and the monotonicity follows. If in addition 
$\|\eta'\|_2\geq C$, then $\lambda_{\Pr}(\Ra, \eta) \leq 0$, by Lemma \ref{lem:existence}, and 
consequently, by \eqref{eq:sting} we have 
\begin{eqnarray}
  \lambda_{\Pra}(\Ra,\eta) \geq   \lambda_{\Pra}(\Ra^*,\eta) + c \min\{1, \Pr\} (\sqrt{\Ra} - \sqrt{\Ra^*}) \,,
\end{eqnarray}
where $c$ depends only on $\mathcal{D}$. 
To prove the continuity, let $(\sqrt{\Ra^*}\u^*, \theta^*)$ be the minimizer of $\lambda_{\Pra}(\Ra,\eta)$. Then, as above with $(\sqrt{\Ra}\overline{\u},\overline{\theta})$ replaced by $(\sqrt{\Ra^*}\u^*, \theta^*)$ we have the following
by the Cauchy inequality
\begin{multline}
  \lambda_{\Pra}(\Ra^*,\eta) 
  \geq \lambda_{\Pra}(\Ra,\eta) + 
  \frac{(\sqrt{\Ra} - \sqrt{\Ra^*})\int_\mathcal{D}\overline{w}\overline{\theta}(\partial_z \eta-1)d\x}{\|\overline{\theta}\|_2^2 + \frac{1}{\Pra}\|\overline{\u}\|_2^2} 
  \\
  \geq \lambda_{\Pra}(\Ra,\eta) -
  \|\eta\|_{W^{1, \infty}([0, 1])} \max\{\Pr, 1\} |\sqrt{\Ra} - \sqrt{\Ra^*}| \,,
\end{multline}
and the lemma follows. 
\end{proof}

We note here that not only is the monotonicity with $\Ra$ of fundamental importance in establishing a rigorous stability result, but it is also invaluable in the algorithmic development.  In simulations we desire to find values of $\Ra$ for which $\lambda_{\Pra}(\Ra,\eta) \sim 0$, and  the monotonicity of $\Ra \mapsto \lambda_{\Pra}(\Ra,\eta)$ allows us to know whether we need to increase or decrease $\Ra$.  The stochastic setting is more complicated,  but the same general principle applies.

Lemma \ref{lem:lambda_monotonic_Ra} immediately yields the following Corollary which we phrase as a remark as it applies only to the deterministic setting:
\begin{remark}
Given $\eta \in C((0,\infty),W^{1,\infty}([0,1]))$, there exists $\Ra_c \in [0,\infty]$ such that
\begin{equation}\label{eq:csth}
\int_0^\infty \lambda_{\Pra}(\Ra,\eta(t)) dt = \infty 
\end{equation}
holds true for all $0\leq \Ra \leq \Ra_c$ and does not hold if $\Ra > \Ra_c$.  
When $\eta$ is replaced by the OU process $\tau^S$, then in general $\Ra_c$ can depend on the realization of the noise. 
However, due to ergodicity, we show 
in the proof of Proposition \ref{prop:stability}
that $\Ra_c$ is almost surely constant. 
\end{remark}

\begin{proof}[Proof of Proposition \ref{prop:stability}]
Exponential decay follows once we establish that
\begin{equation}
\liminf_{t\rightarrow \infty} \frac{1}{t}\int_0^t \lambda_{\Pra}(\Ra,\tau(s))ds > 0.
\end{equation}
Let $\mu_\tau$ be the unique invariant measure for the Ornstein-Uhlenbeck (OU) process $\tau$ defined in \eqref{eq:conductive_time}.  It is known that $\mu_\tau$ is ergodic and supported on the whole state space $\mathcal{Y}$.  Also, by Lemma \ref{lem:lambda_Lip}, the function $\eta \mapsto \lambda_{\Pra}(\Ra,\eta)$ is globally Lipschitz.  In addition, since $\mu_\tau$ is Gaussian it necessarily has $p$-moments for any $p>1$ and we have by standard ergodicity arguments \cite{Sa2017,Ka1997,MeTw2012} that
\begin{equation}
\liminf_{t\rightarrow\infty}\frac{1}{t}\int_0^t\lambda_{\Pra}(\Ra,\tau(s))ds = \int_\mathcal{Y}\lambda_{\Pra}(\Ra,\eta)\mu_\tau(d\eta) = \E \lambda_{\Pra}(\Ra,X),
\end{equation}
where $X$ has law $\mu_\tau$ on $\mathcal{Y}$.

Moreover, by Lemma \ref{lem:lambda_monotonic_Ra}, $\Ra \mapsto \lambda_{\Pra}(\Ra,\eta)$ is decreasing and \eqref{eq:sosmp} holds if $\eta \in K := \{\eta\in \mathcal{Y}: \|\eta'\|_2 \geq c^*\}$ and $c^*$ sufficiently large (see Lemma \ref{lem:lambda_monotonic_Ra}). 
Let $\chi_K$ and $\chi_{K^c}$ be  the characteristic functions of respectively $K$ and the complement $K^c$ of $K$. 
Since $\mu_\tau(K) > 0$, we obtain 
\begin{align}
 \E \lambda_{\Pra}(\Ra^*,\tau^S) &= 
 \E \lambda_{\Pra}(\Ra^*,\tau^S)(\chi_{K} + \chi_{K^c})
 \\
 &\leq \E \lambda_{\Pra}(\Ra^*,\tau^S)(\chi_{K} + \chi_{K^c}) - c(\sqrt{\Ra^*} - \sqrt{\Ra}) \E \chi_{K}   \\
 &= \E \lambda_{\Pra}(\Ra^*,\tau^S) - c(\sqrt{\Ra^*} - \sqrt{\Ra}) \Prb(K)
\end{align}
and \eqref{eq:stexp} follows.  

Finally, since $\tau^S$ has Gaussian distribution on finitely many modes, then $\E \|\tau\|_{W^{1, \infty}([0, 1])} < \infty$, and from \eqref{eq:swcnt} we obtain 
\eqref{eq:idnye}.
\end{proof}

\subsection{Concavity of the the growth factor and implications}\label{sec:rigor_concave}

To determine the functional dependence of $\Ra_c$ on $\gamma$, we first investigate how $\lambda_{\Pr}(\Ra,\zeta(\gamma))$ depends on $\gamma \in\mathbb{R}^M$ the source of the noise in the bulk heating.  Then, we can estimate $\E \lambda_{\Pr}(\Ra,\tau^S)$ by integrating
$\lambda_{\Pr}(\Ra,\zeta(\gamma))$
against the law of $\gamma$ (the law of $\tau^S$) which has a Gaussian distribution. The primary result is Lemma \ref{eq:lem_concave} which establishes the concavity of $\gamma \mapsto \lambda_{\Pr}(\Ra,\zeta(\gamma))$.
Recall the definition of $\zeta$ from \eqref{eq:define_zeta}:
\begin{equation}\label{eq:define_zeta2}
\zeta(\gamma) = \sum_{m=1}^M m\pi \gamma_m \cos(m \pi z),
\end{equation}
so that $\partial_z \tau = \zeta(\gamma)-1$.  Before establishing the concavity  of $\gamma \mapsto \lambda_{\Pr}(\Ra,\zeta(\gamma))$ we show the following deterministic uniform bound on the minimizers of \eqref{eq:lambda_variational_min}.

\begin{lemma}\label{lem:gamma2lambda_bounds}
Let $(\u_\gamma,\theta_\gamma)$ be a minimizer of
\begin{equation}
\lambda_{\Pr}(\Ra,\zeta(\gamma)) = \inf_{(\u,\theta)\in \mathcal{M}} \Big\{\|\nabla \theta\|_2^2 + \frac{1}{\Ra}\|\nabla \u\|_2^2 + \int_{\mathcal{D}}w\theta (\zeta(\gamma)-2)d\x \Big\}.
\end{equation}
If $K$ is any bounded set in $\mathbb{R}^M$ then
\begin{equation}
\sup_{\gamma\in K} \{\|\nabla \theta_\gamma\|_2^2 + \|\nabla \u_\gamma\|_2^2 \} < \infty.
\end{equation}
\end{lemma}

\begin{proof}
Assume that $K \subset B_R$, where $B_R \subset \RR^M$ is the ball of radius $R$ centered at the origin.  Since $(\u_\gamma,\theta_\gamma) \in \mathcal{M}$, then we see that
\begin{align*}
\lambda_{\Pr}(\Ra,\zeta(\gamma)) &\geq \|\nabla\theta_\gamma\|_2^2 + \frac{1}{\Ra}\|\nabla \u_\gamma\|_2^2 - RC\|\u_\gamma\|_2\|\theta_\gamma\|_2\\
&\geq \|\nabla \theta_\gamma\|_2^2 + \frac{1}{\Ra}\|\nabla \u_\gamma\|_2^2 - RC\left(\frac{1}{\Pra\Ra}\|\u_\gamma\|_2^2 + \|\theta_\gamma\|_2^2\right)\\
&\geq \|\nabla\theta_\gamma\|_2^2 + \frac{1}{\Ra}\|\nabla \u_\gamma\|_2^2 - RC,
\end{align*}
where $C$ depends on $\Pra,~\Ra$, and $\|\zeta\|_\infty$ but is independent of $R$ and $\gamma$.  On the other hand, for any fixed $(\u^*,\theta^*)\in\mathcal{M}$,
\begin{align*}
\lambda_{\Pr}(\Ra,\zeta(\gamma)) &\leq \|\nabla\theta^*\|_2^2 + \frac{1}{\Ra}\|\nabla \u^*\|_2^2 + RC\|\u^*\|_2\|\theta^*\|_2\\
&\leq C(R+1).
\end{align*}
Since 
 $\u^*$ and $\theta^*$ are fixed independently of $\gamma$,  the desired result follows.
\end{proof}

 Next, we establish the continuity of the map $\gamma \mapsto \lambda_{\Pr}(\Ra,\zeta(\gamma))$. Although later, we prove a stronger statement (concavity), we need the continuity as a preliminary step.

\begin{Lem}\label{lcnt}
The function $\gamma \mapsto \lambda_{\Pr}(\Ra,\zeta(\gamma))$ is continuous. 
\end{Lem}

\begin{proof}
Fix $\gamma_\infty \in \RR^{M}$ and a sequence $(\gamma_n)_{n = 1}^\infty \subset \RR^{M}$ converging to $\gamma_\infty$. Then
by Lemma \ref{lem:gamma2lambda_bounds}, 
 the sequence of corresponding minimizers  $\{(\bfU_n, \theta_n)\}_{n = 1}^\infty$ 
is bounded in $(H^1)^3 \times H^1$, and therefore it has a weakly convergent subsequence $\{(\bfU_{n_k}, \theta_{n_k})\}_{k = 1}^\infty$ converging to $(\bfU^\ast, \theta^\ast)$. 
By standard imbedding theorems this convergence is strong in $(L^2)^3 \times L^2$ and in particular we have $(\bfU^\ast, \theta^\ast) \in \mathcal M$ and 
\begin{equation}\label{nlcn}
\lim_{k \to \infty}\int w_{n_k} \theta_{n_k} (\zeta(\gamma_{n_k}) - 2) = \int w^\ast \theta^\ast (\zeta(\gamma_\infty) - 2) \,.
\end{equation}
Then by the weak lower semi-continuity of norms we obtain
\begin{equation}\label{eq:wlsc}
\lambda_{\Pr}(\Ra,\zeta(\gamma_\infty))
 \leq \mathcal{Q}_{\Ra}(\u^*,\theta^*,\zeta(\gamma_\infty)) \leq \lim_{k \to \infty} 
 \lambda_{\Pr}(\Ra,\zeta(\gamma_{n_k}))
  \,.
\end{equation}
Since each sequence $(\gamma_n)$ has a convergent subsequence satisfying \eqref{eq:wlsc} we obtain
\begin{equation}\label{eqlif}
\lambda_{\Pr}(\Ra,\zeta(\gamma_\infty)) \leq \liminf_{\gamma \to \gamma_\infty} \lambda_{\Pr}(\Ra,\zeta(\gamma)) \,.
\end{equation}
On the other hand, let $(\bfU_\infty, \theta_\infty) \in \mathcal{M}$
be the minimizer of 
$\lambda_{\Pr}(\Ra,\zeta(\gamma_\infty))$.
Then, for any $n \geq 1$, 
\begin{align}\label{oinq}
\lambda_{\Pr}(\Ra,\zeta(\gamma_{n})) &\leq  
 \|\nabla\theta_\infty\|_2^2+\frac{1}{\ra}\|\nabla\bfU_\infty\|_2^2 + \int 
 w_\infty\theta_\infty (\zeta_(\gamma_\infty) - 2) \\
 &= \lambda_{\Pr}(\Ra,\zeta(\gamma_{\infty})) + \int w_\infty\theta_\infty( \zeta(\gamma_n) - \zeta(\gamma_\infty)) 
 \\
&\leq  \lambda_{\Pr}(\Ra,\zeta(\gamma_{\infty})) + C({\gamma_\infty})|\gamma_\infty - \gamma_n|\,,
\end{align}
and consequently
\begin{equation}\label{eqlsp}
\limsup_{\gamma \to \gamma_\infty} \lambda_{\Pr}(\Ra,\zeta(\gamma)) \leq \lambda_{\Pr}(\Ra,\zeta(\gamma_{\infty}))
\end{equation}
and the assertion follows from \eqref{eqlif} and \eqref{eqlsp}. 
\end{proof}

The continuity established in Lemma \ref{lcnt} allows us to prove the strong convergence of minimizers, which is an importatnt step in the proof of concavity. Specifically, fix a sequence 
$(\gamma_n)_{n = 1}^\infty \subset \RR^M$, such that $\gamma_n \to \gamma_\infty$. 
Let $(\bfU_n, \theta_n)$ be minimizers of $ \lambda_{\Pr}(\Ra,\zeta(\gamma_n))$. By Lemma \ref{lem:gamma2lambda_bounds},  the sequence $\{(\bfU_n, \theta_n)\}_{n = 1}^\infty$ is bounded in
$(H^1)^3 \times H^1$, and therefore up to a sub-sequence  $\{(\bfU_n, \theta_n)\}_{n = 1}^\infty$ is weakly convergent to $(\bfU^*, \theta^*)$
in $(H^1)^3 \times H^1$. The strong convergence is proved in the following corollary.

\begin{Cor}\label{cor:scv}
Fix a sequence 
$(\gamma_n)_{n = 1}^\infty \subset \RR^M$, such that $\gamma_n \to \gamma_\infty$. 
Let $(\bfU_n, \theta_n)$ be the minimizers of $ \lambda_{\Pr}(\Ra,\zeta(\gamma_n))$ and assume 
$\{(\bfU_n, \theta_n)\}_{n = 1}^\infty$ is weakly convergent to
$(\bfU^*, \theta^*)$
in $(H^1)^3 \times H^1$.
 Then $(\bfU^*, \theta^*)$ is a minimizer of $\lambda_{\Pr}(\Ra,\zeta(\gamma_\infty))$ and 
$\{(\bfU_n, \theta_n)\}_{n = 1}^\infty$ converges strongly in $(H^1)^3 \times H^1$  to $(\bfU^*, \theta^*)$.
\end{Cor}

\begin{proof}
By \eqref{eq:wlsc} and the continuity of $\gamma \mapsto \lambda_{\Pr}(\Ra,\zeta(\gamma))$, see Lemma \ref{lcnt}, we have 
\begin{eqnarray*}
 \mathcal{Q}_{\Ra}(\u^*,\theta^*,\zeta(\gamma_\infty)) \leq \lambda_{\Pr}(\Ra,\zeta(\gamma_\infty))
  \,,
\end{eqnarray*}
and therefore $(\bfU^*, \theta^*)$ is a minimizer of $\lambda_{\Pr}(\Ra,\zeta(\gamma_\infty))$. 

From the continuity of 
 $\gamma \mapsto \lambda_{\Pr}(\Ra,\zeta(\gamma))$ and the strong convergence of cross terms (cf. \eqref{nlcn}) we obtain
\begin{equation}\label{eqcof}
 \frac{1}{\ra}\|\nabla \bfU_n\|^2 +  \|\nabla \theta_n\|^2 \to  \frac{1}{\ra}\|\nabla \bfU^*\|^2 +  \|\nabla \theta^*\|^2 \,. 
\end{equation}
In addition, the weak lower semicontinuity of norms yields
\begin{equation}\label{eqcwsc}
    \limsup_{n \to \infty} \|\nabla \bfU_n\|^2 \leq  \|\nabla \bfU^*\|^2 \,,
\qquad 
    \limsup_{n \to \infty} \|\nabla \theta_n\|^2 \leq  \|\nabla \theta^*\|^2
\end{equation}
and by \eqref{eqcof} neither of the inequalities in \eqref{eqcwsc} is strict. Hence, 
\begin{equation}
    \limsup_{n \to \infty} \|\nabla \bfU_n\|^2 =  \|\nabla \bfU^*\|^2 \,,
\qquad 
    \limsup_{n \to \infty} \|\nabla \theta_n\|^2 =  \|\nabla \theta^*\|^2 \,.
\end{equation}
However, weak convergence and convergence of norms in Hilbert spaces imply strong convergence, for example by the use of a parallelogram equality. This completes the proof of the desired assertion. 
\end{proof}

Next, we establish the concavity of $\gamma \mapsto \lambda_{\Pr}(\Ra,\zeta(\gamma))$ and calculate its one-sided derivatives.

\begin{lemma}\label{eq:lem_concave}
The function
$\gamma \mapsto \lambda_{\Pr}(\Ra,\zeta(\gamma))$ is concave and has one sided directional derivatives. Specifically, for any $\overline{\gamma}, \nu \in \mathbb{R}^M$ one has
\begin{align}
\partial_\gamma^+\lambda_{\Pr}(\Ra,\zeta(\overline{\gamma})) &:= \lim_{h\rightarrow 0^+} \frac{\lambda_{\Pr}(\Ra,\zeta((\overline{\gamma}+h\nu))-\lambda_{\Pr}(\Ra,\zeta(\overline{\gamma}))}{h}\\ \label{rhder}
&= \inf_{(\u^*,\theta^*)\in\mathcal{Z}} \int_\mathcal{D} w^*\theta^* \zeta(\nu),\\
\partial_{\gamma}^-\lambda_{\Pr}(\Ra,\zeta(\overline{\gamma})) &:= \lim_{h\rightarrow 0^-}\frac{\lambda_{\Pr}(\Ra,\zeta((\overline{\gamma}+h\nu))-\lambda_{\Pr}(\Ra,\zeta(\overline{\gamma}))}{h}\\ \label{lhder}
&= \sup_{(\u^*,\theta^*)\in\mathcal{Z}}\int_{\mathcal{D}}w^*\theta^*\zeta(\nu),
\end{align}
where $\mathcal{Z}$ is the set of all global minimizers of $\lambda_{\Pr}(\Ra,\zeta(\overline{\gamma}))$.
\end{lemma}

\begin{proof}

Fix any sequence $\gamma_n \to \overline{\gamma}$ and any 
$(\u_{\bar{\gamma}},\theta_{\bar{\gamma}}) \in \mathcal{Z}$.  Since $\lambda_{\Pr}(\Ra,\zeta(\gamma_n))$ is the minimum over $\mathcal{M}$ and $(\u_{\bar{\gamma}},\theta_{\bar{\gamma}}) \in \mathcal{M}$, then
\begin{align}\label{eq:gamma2lambda_cont1}
\lambda_{\Pr}(\Ra,\zeta(\gamma_n)) &\leq \|\nabla\theta_{\bar{\gamma}}\|_2^2 + \frac{1}{\Ra}\|\nabla\u_{\bar{\gamma}}\|_2^2 + \int_\mathcal{D}w_{\bar{\gamma}}\theta_{\bar{\gamma}}(\zeta(\gamma_n)-2)d\x\\ \label{eq:gamma2lambda_cont2}
&= \lambda_{\Pr}(\Ra,\zeta(\bar{\gamma})) + \int_\mathcal{D} w_{\bar{\gamma}}\theta_{\bar{\gamma}}(\zeta(\gamma_n)-\zeta(\bar{\gamma}))d\x \,.
\end{align}
Since $(\u_{\bar{\gamma}},\theta_{\bar{\gamma}}) \in \mathcal{Z}$ was arbitrary, by setting $\gamma_n := \overline{\gamma}+h_n\nu$ for some $h_n \to 0$, we obtain
by linearity of $\gamma \mapsto \zeta(\gamma)$ that
\begin{equation}\label{simc}
\begin{aligned}
 \lambda_{\Pr}(\Ra,\zeta(\overline{\gamma}+h_n\nu)) &- \lambda_{\Pr}(\Ra,\zeta(\overline{\gamma})) \\ &\leq \inf_{(\u^*,\theta^*)\in\mathcal{Z}_\infty} \int_\mathcal{D} w^*\theta^*(\zeta(\overline{\gamma}+h_n\nu) - \zeta(\overline{\gamma}))d\x\\
&= h_n \inf_{(\u^*,\theta^*)\in\mathcal{Z}_\infty} \int_\mathcal{D}w^*\theta^* \zeta(\nu)\,.
\end{aligned}
\end{equation}
On the other hand, for any minimizer $(\u_n, \theta_n)$ of 
$\lambda_{\Pr}(\Ra,\zeta(\overline{\gamma}))$ we have 
\begin{align*}
\lambda_{\Pr}(\Ra,\zeta(\overline{\gamma})) &\leq \|\nabla \theta_n\|_2^2 + \frac{1}{\Ra}\|\nabla\u_n\|_2^2 + \int_\mathcal{D}w_n\theta_n(\zeta(\overline{\gamma})-2)d\x\\
&= \lambda_{\Pr}(\Ra,\zeta(\overline{\gamma}+h_n\nu)) + \int_\mathcal{D} w_n\theta_n(\zeta(\overline{\gamma})-\zeta(\overline{\gamma}+h_n\nu))d\x,
\end{align*}
and consequently by linearity of $\gamma \mapsto \zeta(\gamma)$
\begin{equation}\label{aies}
\lambda_{\Pr}(\Ra,\zeta(\overline{\gamma}+h_n\nu)) - \lambda_{\Pr}(\Ra,\zeta(\overline{\gamma})) \geq h_n \int_\mathcal{D} w_n \theta_n \zeta(\nu).
\end{equation}
To prove \eqref{rhder}, if $h_n > 0$, then by \eqref{simc} and \eqref{aies}
\begin{align*}
\int_\mathcal{D}w_n\theta_n\zeta(\nu) d\x &\leq \frac{\lambda_{\Pr}(\Ra,\zeta(\overline{\gamma}+h_n\nu)) - \lambda_{\Pr}(\Ra,\zeta(\overline{\gamma}))}{h_n}\\
&\leq \inf_{(\u^*,\theta^*)\in\mathcal{Z}_\infty} \int_\mathcal{D} w^*\theta^*\zeta(\nu)d\x.
\end{align*}
Since by Corolloary \ref{cor:scv} the sequence
$\{(\overline{\u}_n,\overline{\theta}_n)\}_{n=1}^\infty$ converges strongly in $(H^1)^3\times H^1$ to some $(\u_\infty,\theta_\infty)\in\mathcal{Z}_\infty$ we obtain
\begin{align*}
\int_\mathcal{D}w_\infty\theta_\infty \zeta(\nu) d\x &\leq \lim_{n\rightarrow \infty}\frac{\lambda_{\Pr}(\Ra,\zeta(\overline{\gamma}+h_n\nu))-\lambda_{\Pr}(\Ra,\zeta(\overline{\gamma}))}{h_n}\\
&\leq \inf_{(\u^*,\theta^*)\in\mathcal{Z}_\infty} \int_\mathcal{D}w^*\theta^* \zeta(\nu)d\x \,.
\end{align*}
Since  the sequence $(h_n)\subset (0,\infty)$ was arbitrary, 
\begin{equation}
\partial_\nu^+\lambda_{\Pr}(\Ra,\zeta(\overline{\gamma})) = \inf_{(\u^*,\theta^*)\in\mathcal{Z}_\infty}\int_\mathcal{D}w^*\theta^*\zeta(\nu) d\x.
\end{equation}

To prove \eqref{lhder} assume $h_n < 0$, which by \eqref{simc} and \eqref{aies} implies
\begin{align*}
\int_\mathcal{D} w_n\theta_n\zeta(\nu) d\x &\geq \frac{\lambda_{\Pr}(\Ra,\zeta(\overline{\gamma}+h_n\nu))-\lambda_{\Pr}(\Ra,\zeta(\overline{\gamma}))}{h_n}\\
&\geq \sup_{(\u^*,\theta^*)\in\mathcal{Z}_\infty}\int_\mathcal{D}w^*\theta^*\zeta(\nu)d\x.
\end{align*}
Employing strong convergence just as above, we find that:
\begin{equation}
\partial_\nu^-\lambda_{\Pr}(\Ra,\zeta(\overline{\gamma})) = \sup_{(\u^*,\theta^*)\in\mathcal{Z}_\infty}\int_\mathcal{D}w^*\theta^*\zeta(\nu) d\x.
\end{equation}

Finally, concavity of $\gamma \mapsto \lambda_{\Pr}(\Ra,\zeta(\gamma))$ is equivalent to showing that for any $\gamma_1,\gamma_2\in\mathbb{R}^M$ and $t\in (0,1)$ one has 
\begin{align*}
0 &\leq \lambda_{\Pr}(\Ra,\zeta(t\gamma_1 + (1-t)\gamma_2)) - t\lambda_{\Pr}(\Ra,\zeta(\gamma_1))-(1-t)\lambda_{\Pr}(\Ra,\zeta(\gamma_2))\\
&= t\left[\lambda_{\Pr}(\Ra,\zeta(\gamma_1+(1-t)(\gamma_2-\gamma_1)))-\lambda_{\Pr}(\Ra,\zeta(\gamma_1))\right]\\
&+(1-t)\left[\lambda_{\Pr}(\Ra,\zeta(\gamma_2+t(\gamma_1-\gamma_2)))-\lambda_{\Pr}(\Ra,\zeta(\gamma_2))\right].
\end{align*}
After dividing by $t(1-t)>0$, this is equivalent to showing that
\begin{align}\nonumber
&\frac{\lambda_{\Pr}(\Ra,\zeta(\gamma_1+(1-t)\nu))-\lambda_{\Pr}(\Ra,\zeta(\gamma_1))}{1-t}\\&+\frac{\lambda_{\Pr}(\Ra,\zeta(\gamma_2-t\nu))-\lambda_{\Pr}(\Ra,\zeta(\gamma_2))}{t}\geq 0,
\end{align}
where $\nu=\gamma_2-\gamma_1$.  However, by 
\eqref{rhder}, \eqref{lhder}, and
$\gamma_1+(1-t)\nu = \gamma_2-t\nu$ we see that
\begin{align*}
& \frac{\lambda_{\Pr}(\Ra,\zeta(\gamma_1+(1-t)\nu))-\lambda_{\Pr}(\Ra,\zeta(\gamma_1))}{1-t} - \frac{\lambda_{\Pr}(\Ra,\zeta(\gamma_2-t\nu))-\lambda_{\Pr}(\Ra,\zeta(\gamma_2))}{-t}\\
 &\geq \int_\mathcal{D}w_{\gamma_1+(1-t)\nu}\theta_{\gamma_1+(1-t)\nu}\zeta(\nu) d\x - \int_\mathcal{D} w_{\gamma_2-t\nu}\theta_{\gamma_2-t\nu}\zeta(\nu) d\x\\
&= 0,
\end{align*}
and the concavity follows.
\end{proof}

The following bound on the one-sided derivatives is used when 
we integrate against a Gaussian random variable below. 

\begin{corollary}\label{cor:der_bounds}
For any $\gamma \in \mathbb{R}^M$,
\begin{equation}
\left|\partial_\gamma^{\pm}\lambda_{\Pr}(\Ra,\zeta(\gamma))\right| \leq \frac{\sqrt{\Ra\Pr}}{2\pi} \sum_{k=1}^M \frac{|\gamma_k|}{k}.
\end{equation}
\end{corollary}

\begin{proof}
Using the definition of $\zeta$ from \eqref{eq:define_zeta}, and the definition of $\mathcal{M}$ from \eqref{eq:variational_constraint}, we have for any $(\u,\theta)\in\mathcal{M}$,
\begin{align}
\left|\int_\mathcal{D} w \theta \zeta(\gamma)d\x\right| &\leq \frac{\sqrt{\Ra\Pr}}{2}\|\zeta(\gamma)\|_\infty\left(\|\theta\|_2^2 + \frac{1}{\Pr\Ra}\|\u\|_2^2\right)\\
&\leq \frac{\sqrt{\Ra\Pr}}{2\pi}\sum_{k=1}^M \frac{|\gamma_k|}{k}.
\end{align}
The infimum and supremum preserve this inequality and the desired result follows from Lemma \ref{eq:lem_concave}.
\end{proof}

\subsection{Shape of the growth factor and consequences}\label{sec:rigor_find_gammac}
In this section we establish estimates on the   growth factor $\lambda$ and obtain bounds on the critical Rayleigh number as a function of the strength of the forcing $H$. 

\subsubsection{Basic estimates on the growth factor}
Recall that in Lemma \ref{eq:lem_concave} we established that $\gamma \mapsto \lambda_{\Pr}(\Ra,\zeta(\gamma))$ is concave, and we derived bounds on its one-sided derivatives.  Next, we need to understand the behavior at $\gamma=0$.

If $\gamma=0$, then
\begin{equation}
\lambda_{\Pr}(\Ra,\zeta(0)) = \inf_{(\u,\theta)\in\mathcal{M}} \frac{1}{\Ra}\|\nabla\u\|_2^2 + \|\nabla \theta\|_2^2 - 2\int_\mathcal{D} w\theta d\x.
\end{equation}
Using standard arguments from the Calculus of Variations \cite{DoGi1995}, we obtain that 
$w$ and $\theta$ satisfy
\eqref{eq:four-w} and \eqref{eq:four-th} with $\partial_z \tau = -1$ and recall that $\u$ satisfies
 stress-free boundary conditions \eqref{bc:str-free}. As in \cite{Ra1916} we can expand $w$
into Fourier  series 
\begin{eqnarray}
 w(x', z) = \sum_{k \in \mathbb{Z}^2_L}\sum_{m \in \mathbb{Z}_2} \hat{w}_{k, m} \xi_k(x') \sin(m\pi z) \,,   
\end{eqnarray}
where $x = (x' , z) \in \RR^2\times \RR$ and 
\begin{eqnarray}
    \xi_k(x') = 
    \begin{cases}
        \cos(k\cdot x') & k_2 > 0 \textrm{ or } k_2 = 0 \textrm{ and } k_1 > 0 \\
        \sin(k\cdot x') & \textrm{ otherwise } \,.
    \end{cases}
\end{eqnarray}
We remark that for the no-slip boundary conditions, we have to use  expansions with less explicit eigenfunctions that lead to rather complicated expressions. 
Thus, to avoid unnecessary and tedious manipulations, we opted to only discuss stres-free boundary conditions. 
Then, \eqref{eq:four-w} and \eqref{eq:four-th} with $\partial_z \tau = -1$ becomes
\begin{align}\label{eq:hsys-w}
\frac{\lambda}{\Pra\Ra}\left(m^2+|k|^2\right) \hat{w}_{k, m} &= \frac{1}{\Ra}\left(m^2 + |k|^2\right)^2\hat{w}_{k,m} - |k|^2 \hat{\theta}_{k,m} \,,\\\label{eq:hsys-th}
\lambda \hat{\theta}_{k,m} &= - \hat{w}_{k,m} + \left(m^2+ |k|^2\right)\hat{\theta}_{k,m}  \,.
\end{align}
Moreover, for given $m \in \mathbb{Z}_2 \setminus \{0\}$ and $k \in \mathbb{Z}^2_L $, \eqref{eq:hsys-w}, \eqref{eq:hsys-th} has a non-trivial solution $(\hat{w}_{k, m}, \hat{\theta}_{k,m}) \neq (0, 0)$ if and only if (zero determinant and quadratic equation):
\begin{multline}\label{eq:lakm}
\lambda_{\Pr,k,m}(\Ra,\zeta(0)) \\
= \frac{1}{2}\left((\Pr+1)(m^2+|k|^2) - \sqrt{(\Pr-1)^2(m^2+|k|^2)^2 + \frac{4\Pr\Ra}{m^2+|k|^2}|k|^2}\right),
\end{multline}
where we have chosen the negative sign in the quadratic root in \eqref{eq:lakm} as we are interested in the smallest eigenvalue $\lambda_{\Pr,k,m}(\Ra,\zeta(0))$ (see the discussion below \eqref{eq:v-eqn}--\eqref{eq:th-eqn}).  To find the minimum with respect to $k$ and $m$ we first compute for each fixed 
 $m \in \mathbb{Z}_2 \setminus \{0\}$ and $k \in \mathbb{Z}^2_L$
 we have 
\begin{align*}
\frac{\partial \lambda_{\Pr,k,m}(\Ra,\zeta(0))}{\partial m} &= m\left(\Pr + 1 - \frac{(\Pr-1)^2(m^2+|k|^2) - \frac{2\Pr \Ra}{(m^2+|k|^2)^2}|k|^2}{\sqrt{(\Pr-1)^2(m^2+|k|^2) + \frac{4\Pr\Ra}{m^2+|k|^2}|k|^2}}\right)\\
&\geq m\left(\Pr+1 - \frac{(\Pr-1)^2(m^2+|k|^2)}{\sqrt{(\Pr-1)^2(m^2+|k|^2)^2}}\right)\\
&= 2m > 0.
\end{align*} 
Thus minimum $m \in \mathbb{Z}_2 \mapsto \lambda_{\Pr,k,m}(\Ra,\zeta(0))$ occurs for the minimal vertical wave number $m=\pi$.  The minimum of $k \mapsto \lambda_{\Pr, k, m}(\Ra,\zeta(0))$ is slightly more complicated and is analyzed in Lemma 
\ref{lem:eoran} below. Note that $\lambda_{\Pr, k, m}(\Ra,\zeta(0))$ depends only on $|k|$ rather than on $k$.

Recall that for each $\Ra$ and $\zeta$
\begin{eqnarray}
   \lambda_{\Pr}(\Ra,\zeta) = \inf_{k \in \mathbb{Z}^2_L, m \in \mathbb{Z}_2}  
   \lambda_{\Pr, k, m}(\Ra,\zeta) \,.
\end{eqnarray}
Next, we show in Lemma \ref{lem:gamma_0_bounds}  below that if  $\lambda_{\Pr}(\Ra,\zeta(0)) < 0$ for some $\Ra$, then $\lambda_{\Pr}(\Ra,\zeta(\gamma)) < 0$ for all $\gamma$ and hence by \eqref{eq:tauS_ergodicity} the conductive state is trivially unstable.
 The following lemma gives the optimal upper bound on the Rayleigh number which guarantees that $\lambda_{\Pr}(\Ra,\zeta(0))>0$.

\begin{lemma}\label{lem:eoran}
To simplify calculations, assume that $\Pr=1$, and define 
\begin{align}
 j_0 &= \sup\Big\{\sqrt{a^2 + b^2} \leq \frac{L}{2\sqrt{2}}: a, b \in \mathbb{Z} \Big\} \,, \\
 j_1 &= \inf \Big\{\sqrt{a^2 + b^2} \geq \frac{L}{2\sqrt{2}}: a, b \in \mathbb{Z} \Big\} \,.
\end{align}
 Also denote $r_0 = \frac{2}{L}j_0$ and $r_1 = \frac{2}{L}j_1$ and observe that for large $L$, one has $r_0,~r_1 \approx \frac{1}{\sqrt{2}}$, and $L<2\sqrt{2}$ for $j_0 = 0$.  Then, $\min_{k \in \mathbb{Z}^2_L}\lambda_{\Pr=1, k, m = \pi}(\Ra,\zeta(0)) \geq 0$ if and only if
\begin{equation}
\Ra \leq \pi^4 \min_{l \in \{0,1\}} \frac{(1+r_l^2)^3}{r_l^2} \,,
\end{equation}
where we set $\frac{(1+r_l^2)^3}{r_l^2} := \infty$
 if $r_l = 0$. 
\end{lemma}

\begin{proof}
From \eqref{eq:lakm} and the fact that the minimum is attained at $m = \pi$,  $\lambda_{\Pr=1, k, m = \pi}(\Ra,\zeta(0))\geq 0$ is equivalent to
\begin{equation}
2(\pi^2+|k|^2) \geq \sqrt{\frac{4\Ra |k|^2}{\pi^2+|k|^2}},
\end{equation}
for all $k \in \mathbb{Z}^2_L$, and therefore 
\begin{equation}\label{eq:Ra_critical_m_k}
\Ra \leq \frac{(\pi^2+|k|^2)^3}{|k|^2},
\end{equation}
for all $k \in \mathbb{Z}^2_L$.
The right hand side of \eqref{eq:Ra_critical_m_k} is  
a convex function of $|k|$ and its
minimum is attained for $|k|^2=\frac{\pi^2}{2}$. 
After substitution $|k|^2 = \frac{4\pi^2 |j|^2}{L^2}$ we obtain that the unique minimum (even unique critical point) is attained at $|j|=\frac{L}{2\sqrt{2}}$. By the definition of $\mathbb{Z}^2_L$, we have that $j$ is a vector with integer coefficients, and therefore the minimum is attained 
either at $j_0$ or $j_1$.
A substitution into \eqref{eq:Ra_critical_m_k} gives the desired result.
\end{proof}

\begin{remark}
We want to emphasize that the restriction $\Pr=1$ is not necessary for validity of Lemma \ref{lem:eoran}, but reduces algebraic manipulations in the proof and  very similar results hold for all $\Pr>0$.

Also, in almost all of the derivations below we will assume that $L < 2\sqrt{2}$ to simplify the definition of $j_0$ and $j_1$ in Lemma \ref{lem:eoran}, which become $j_0 = 0$ and $j_1=1$.  This isn't an essential hypothesis, but it will let us avoid number theoretical discussion and make the relevant calculations more tractable.
\end{remark}

The following result gives $\Ra$ dependent estimates on the growth factor $\lambda$ at and near $\gamma=0$. In other words, we estimate the behavior of the growth factor when the internal heating is very small. 

\begin{lemma}\label{lem:gamma_0_bounds}
Assume that $\Pr=1,~ L\leq 2\sqrt{2},~2\in\mathcal{N}$, and $\lambda_{\Pr}(\Ra,\zeta(0)) >0$ (or equivalently $\Ra \leq \pi^4\frac{(4+L^2)^3}{4L^4}$, see Lemma \ref{lem:eoran}).  Then we have the following results dependent on the value of $\Ra$:
\begin{enumerate}
\item If $\Ra \leq \frac{4\pi^4(4+L^2)}{L^4}$, then
\begin{equation}
\lambda_{\Pr=1}(\Ra,\zeta(0)) = \pi^2,\quad\mbox{ and } \quad \partial_\gamma^+\lambda_{\Pr=1}(\Ra,\zeta(\gamma=0)) = 0.
\end{equation}
\item If $\frac{4\pi^4(4+L^2)}{L^4} \leq \Ra \leq \frac{\pi^4(4+L^2)^3}{4L^4}$, then:
\begin{align}
\lambda_{\Pr=1}(\Ra,\zeta(0)) &= \pi^2\left(1+\frac{4}{L^2}\right) - \frac{2\sqrt{\Ra}}{\sqrt{4+L^2}},\\
\partial_\gamma^+\lambda_{\Pr=1}(\Ra,\zeta(\gamma=0)) &= -\frac{\sqrt{\Ra} \gamma_2}{2\pi\sqrt{4+L^2}}.
\end{align}
We remark that if $2\not\in \mathcal{N}$, then $\partial_\gamma^+\lambda_{\Pr=1}(\Ra,\zeta(\gamma=0)) = 0$.
\item Finally, if $\Ra > \frac{\pi^4(4+L^2)^3}{4L^4}$, then $\lambda_{\Pr=1}(\Ra,\zeta(\gamma)) \leq 0$ for all $\gamma \in \mathbb{R}^\mathcal{N}$.
\end{enumerate}
\end{lemma}

\begin{proof}
First we prove part 3 assuming that part 2 holds. Indeed, by part 2 for $\Ra_U := \frac{\pi^4(4+L^2)^3}{4L^4}$, then $\lambda_{\Pr=1}(\Ra_U,\zeta(0)) = 0$
and $\partial_\gamma^+\lambda_{\Pr=1}(\Ra_U,\zeta(\gamma=0)) \leq 0$, and therefore by concavity $\lambda_{\Pr=1}(\Ra_U,\zeta(\gamma)) \leq 0$ for each $\gamma$. Moreover, by Lemma \ref{lem:lambda_monotonic_Ra} the function 
$\Ra \mapsto \lambda_{\Pr=1}(\Ra,\zeta(\gamma))$
is non-increasing, and thus $\lambda_{\Pr=1}(\Ra,\zeta(\gamma)) \leq 0$ for each $\Ra \geq \Ra_U$, as desired.

To prove parts 1 and 2, we assume 
$\Ra < \Ra_U$ note that the minimum with respect to $m$ is attained at $m = \pi$ and $\lambda$ depends only on $|k|$.
To find the minimum of the function $|k| \mapsto \lambda_{\Pr=1, m = \pi}(\Ra,\gamma(0))$ we calculate
\begin{equation*}
\frac{\partial \lambda_{\Pr=1,k,m = \pi}(\Ra,\zeta(0))}{\partial |k|} = |k|\left(2-\frac{\sqrt{\Ra}}{(\pi^2+|k|^2)^{3/2}}\frac{\pi^2}{|k|}\right) \,,
\end{equation*}
and observe that if 
$|k| = 0$, then the derivative is negative, and therefore $|k| = 0$ is not the minimizer.
Using $\Ra < \Ra_U$, we see that 
\begin{equation*}
2 - \frac{\sqrt{\Ra}}{(\pi^2+|k|^2)^{3/2}}\frac{\pi^2}{|k|} \geq 2 - \frac{\pi^2}{|k|^2} > 0
\end{equation*}
whenever $|k|^2>\pi^2/2$.  Thus, $k \mapsto \lambda_{\Pr=1,k,m = \pi}(\Ra,\zeta(0))$ is an increasing function of $|k|$ for $|k|\geq \pi/\sqrt{2}$. Since $|k| = \frac{2\pi}{L}|j|$ for some $j \in \mathbb{Z}^2$ and $L \leq 2\sqrt{2}$, 
then the function increases if $|j| \geq 1$.

Thus the minimum of $k \mapsto \lambda_{\Pr=1,k,m = \pi}(\Ra,\zeta(0))$ is attained either at $|j| = 0$
or $|j| = 1$, that is, at $k=0$ or $k=\frac{2\pi}{L}$.

This leads to two possible cases if \eqref{eq:Ra_critical_m_k} is satisfied:
\begin{align*}
\lambda_{\Pr=1,k=0,m=\pi}(\Ra,\zeta(0)) &= \pi^2,\\
\lambda_{\Pr=1,k=2\pi/L,m=\pi}(\Ra,\zeta(0)) &= \pi^2\left(1+\frac{4}{L^2} - \sqrt{\frac{4\Ra}{\pi^4(L^2+4)}}\right).
\end{align*}
After standard algebraic manipulation we find that $\lambda_{\Pr=1,k=2\pi/L,m=\pi}(\Ra,\zeta(0)) \leq \lambda_{\Pr=1,k=0,m=\pi}(\Ra,\zeta(0))$ if
and only if 
\begin{equation}
\Ra \leq \frac{4\pi^4(4+L^2)}{L^4}.
\end{equation}
With this in mind we are ready to analyze parts 1 and 2 of the lemma.

\begin{enumerate}
\item If $\Ra \leq \frac{4\pi^4(4+L^2)}{L^4}$, then  $\min_{k} \lambda_{\Pr=1, k, m = \pi}(\Ra,\zeta(0)) = \pi^2$ is attained at $k = 0$.  Since $m = \pi$ and $k = 0$, then a substitution into \eqref{eq:hsys-w} and \eqref{eq:hsys-th} and standard manipulations imply that the minimum is attained at $\u = 0$ and $\theta = \sin (\pi z)$. We chose the normalization such that $(\u, \theta) \in \mathcal{M}$, that is,  $\|\theta\|^2_2 = 1$.  Then Lemma \ref{eq:lem_concave} implies $\partial_\gamma \lambda_{\Pr=1}(\Ra,\gamma(0)) = 0$.
\item If $\frac{4\pi^4(4+L^2)}{L^4} \leq \Ra \leq \frac{\pi^4(4+L^2)^3}{4L^4}$, then $m = \pi$ and 
$|k| = \frac{2\pi}{L}$ for which
\begin{equation}
\lambda_{\Pr=1}(\Ra,\zeta(0)) = \pi^2\left(1+\frac{4}{L^2}\right) - \frac{2\sqrt{\Ra}}{\sqrt{4+L^2}}
\end{equation}
and the the minimum in the defintion of $\lambda$
is achieved for 
\begin{align*}
\theta &= \frac{\sqrt{2}}{L}\sin(\pi z)\sin(2\pi x/L),\quad w = \frac{2\sqrt{2\Ra}}{L\sqrt{4+L^2}}\sin(\pi z)\sin(2\pi x/L),\\
u &= \frac{\sqrt{2\Ra}}{\sqrt{4+L^2}}\cos(\pi z)\cos(2\pi x/L),\quad v = 0 \,.
\end{align*}
where the velocity field is $\u(x,y,z) = (u,v,w)^T$.
We remark that there are other minimizers shifted by a phase in $x$-direction or with $x$ and $y$-directions exchanged, but they give the same results. 
The normalization constants are chosen such that the normalization  $1 = \|\theta\|^2 + \frac{1}{\Ra}\|\u\|^2$ is satisfied and the velocity field is indeed incompressible. Then, by Lemma \ref{eq:lem_concave}  we calculate the derivative
\begin{align*}
\partial_\gamma^+\lambda_{\Pr=1}(\Ra,\zeta(0)) &= \int_{\mathcal{D}} w\theta \zeta(\gamma)\\
&= \frac{2\sqrt{\Ra}}{\sqrt{4+L^2}}\int_0^1 \sin^2(\pi z) \zeta(\gamma) dz\\
&= - \frac{2\sqrt{\Ra}}{\sqrt{4+L^2}}\int_0^1 \cos(2\pi z) \zeta(\gamma) dz\\
&= - \frac{\sqrt{\Ra}}{2\pi\sqrt{4+L^2}}\gamma_2,
\end{align*} 
where $\gamma_2$ is the coefficient for $m=2$ in \eqref{eq:define_zeta2}.
If $2\not\in\mathcal{N}$, then $\zeta(\gamma)$ is orthogonal to the test functions $w\theta$ in $L^2$ so the derivative vanishes.
% \item For $\Ra > \frac{\pi^4(4+L^2)^3}{4L^4} = \Ra_L$ we first note that $\lambda_{\Pr=1}(\Ra_L,\zeta(0)) = 0$, and $\partial_\gamma^+\lambda_{\Pr=1}(\Ra_L,\zeta(0)) \leq 0$.  Concavity of $\lambda_{\Pr=1}(\Ra,\zeta(\gamma))$ with respect to $\gamma$ then indicates that $\lambda_{\Pr=1}(\Ra_L,\zeta(\gamma)) \leq 0$ for all $\gamma$, and hence monotonicity of $\lambda$ with respect to $\Ra$ ensures that $\lambda_{\Pr=1}(\Ra,\zeta(\gamma))\leq 0$ for all $\Ra>\Ra_L$.
\end{enumerate}

\end{proof}

\subsubsection{Asymptotic behavior of $\Ra_c(H)$}\label{sec:rigor_H_dep}
In all of the following we will make the simplifying assumption that $\mathcal{N} = \{2\}$, that is, only the $2$nd mode is forced.  This restriction is not necessary, but significantly reduces the calculations that follow.  We do want to emphasize  that forcing the $2$nd mode does seem necessary for the results that follow, but we can not determine if this is only a product of our method of proof, or if this is in fact a physical property of the system.

\begin{proof}[Proof for Theorem \ref{thm:lower_bounds}]
Fix any $\overline{\Ra} \in [\Ra_L, \Ra_U)$. Then
from Lemma \ref{lem:gamma_0_bounds} it follows that 
$\lambda_{\Pr=1}(\overline{\Ra}, 0) > 0$ and 
$\partial_\gamma\lambda_{\Pr=1}(\overline{\Ra}, 0) < 0$, and consequently by concavity and continuity
proved in Lemma \ref{eq:lem_concave} 
we have that $\gamma \mapsto \lambda_{\Pr=1}(\Ra, \zeta(\gamma))$ is strictly decreasing.
 In particular,  
 there is exactly one value $\overline{\gamma}$ so that $\lambda_{\Pr=1}(\Ra_L,\zeta(\overline{\gamma})) = 0$.

The concavity of the growth factor $\lambda$ yields the following bounds (see Figure \ref{fig:lambda_1})
for each $\gamma > 0$:
\begin{align} \label{eq:bogb}
&\lambda_{\Pr=1}(\overline{\Ra},\zeta(0)) - \gamma \sup_{\gamma>0}\left| \partial_\gamma^+\lambda_{\Pr=1}(\overline{\Ra},\zeta(\gamma))\right| \\
&\leq \lambda_{\Pr=1}(\overline{\Ra},\zeta(\gamma)) \\
&\leq \lambda_{\Pr=1}(\overline{\Ra},\zeta(0)) - \gamma \left| \partial_\gamma^+\lambda_{\Pr=1}(\overline{\Ra},\zeta(0))\right|.
\end{align}
Because $\lambda_{\Pr=1}(\overline{\Ra},\zeta(\overline{\gamma})) = 0$, then
\begin{equation}
\frac{\lambda_{\Pr=1}(\overline{\Ra},\zeta(0))}{|\sup_{\gamma>0}\left|\partial_\gamma^+\lambda_{\Pr=1}(\overline{\Ra},\zeta(\gamma))\right|} \leq \overline{\gamma} \leq \frac{\lambda_{\Pr=1}(\overline{\Ra},\zeta(0))}{\left|\partial_\gamma^+\lambda_{\Pr=1}(\overline{\Ra},\zeta(0))\right|} \,,
\end{equation}
that is, given $\overline{\Ra}$, we obtained bounds on the strength of the deterministic internal heating that yields marginal criticality. 

The concavity and \eqref{eq:bogb} also yields the lower bound (see  Figure \ref{fig:lambda_2})
\begin{align}\label{eq:lambda_lower_bound}
&\lambda_{\Pr=1}(\overline{\Ra},\zeta(\gamma))\\
& \geq \begin{cases}  (\overline{\gamma} - \gamma)\frac{\lambda_{\Pr=1}(\overline{\Ra},\zeta(0))}{\overline{\gamma}} \geq (\overline{\gamma} - \gamma)\left|\partial_\gamma^+\lambda_{\Pr=1}(\overline{\Ra},\zeta(0))\right| & \gamma \in (0,\overline{\gamma}),\\
(\overline{\gamma} - \gamma)\sup_{\gamma>0}\left|\partial_\gamma^+\lambda_{\Pr=1}(\overline{\Ra},\zeta(\gamma))\right| & \gamma \geq \overline{\gamma}\end{cases}\nonumber
\end{align}
Using the hypotheses that $\mathcal{N} = \{2\}$, 
$\overline{\Ra} \in [\Ra_L, \Ra_U)$, and $L \leq 2\sqrt{2}$ we see from Lemma \ref{lem:gamma_0_bounds} and Corollary \ref{cor:der_bounds} that
\begin{equation}\label{eq:lambda_ratio_bound}
\frac{\sup_{\gamma>0}\left|\partial_\gamma^+\lambda_{\Pr=1}(\overline{\Ra},\zeta(\gamma))\right|}{\left|\partial_\gamma^+\lambda_{\Pr=1}(\overline{\Ra},\zeta(0))\right|} \leq \frac{\sqrt{4+L^2}}{2} \leq \sqrt{3}.
\end{equation}

Now we estimate $\E_{\gamma}[\lambda_{\Pr=1}(\overline{\Ra},\zeta(\gamma)]$, where $\gamma = N_H \sim \mathcal{N}\left(0,\frac{\gamma_H^2}{8\pi^2}\right)$ (recall that we are only considering $k=2$). Note that after the substitution, $z \to 1-z$, we obtain that $\zeta(\gamma) \to -\zeta(\gamma)$, andtherefore it suffices to assume $\gamma \geq 0$. In particular, we bound 
$\E_{\gamma}[\lambda_{\Pr=1}(\overline{\Ra},\zeta(\gamma)]$, where $\gamma = |N_H|$. 

Let $C_{\gamma_H}$ be the relevant normalization constant of the normal distribution.  Then
using \eqref{eq:lambda_lower_bound} and \eqref{eq:lambda_ratio_bound} we see that
\begin{align}\nonumber
&\frac{C_{\gamma_H}}{\left|\partial_\gamma^+\lambda_{\Pr=1}(\Ra^*,\zeta(0))\right|} \E_\gamma[\lambda_{\Pr=1}(\Ra^*,\zeta(\gamma))]\\ &\geq \int_0^{\gamma_H}(\gamma_H-\gamma)e^{-\frac{\gamma^2}{2\gamma_H^2/(8\pi^2)}}d\gamma
+ \sqrt{3}\int_{\gamma_H}^\infty (\gamma_H - \gamma)e^{-\frac{\gamma^2}{2\gamma_H^2/(8\pi^2)}}d\gamma.
\end{align}
Using the change of variables $\gamma = x\frac{\gamma_H}{2\pi\sqrt{2}}$ we have
\begin{align*}
&\frac{2\pi\sqrt{2}C_{\gamma_H}}{\left|\partial_\gamma^+\lambda_{\Pr=1}(\Ra^*,\zeta(0))\right|\gamma_H^2}\E_\gamma[\lambda_{\Pr=1}(\Ra^*,\zeta(\gamma))]\\ &\geq \int_0^{2\pi\sqrt{2}}\left(1-\frac{x}{2\pi\sqrt{2}}\right)e^{-\frac{x^2}{2}}dx 
+ \sqrt{3}\int_{2\pi\sqrt{2}}^\infty\left(1-\frac{x}{2\pi\sqrt{2}}\right)e^{-\frac{x^2}{2}}dx\\
&\geq \int_0^{\infty}e^{-\frac{x^2}{2}}dx - \frac{\sqrt{3}}{2\pi\sqrt{2}}\int_{0}^\infty xe^{-\frac{x^2}{2}}dx = \frac{\sqrt{2\pi}}{2} -  \frac{\sqrt{3}}{2\pi\sqrt{2}}> 0,
\end{align*}
and the result follows. 
\end{proof}

\subsubsection{Gaussian tails for the growth factor}\label{sec:rigor_tails}
To prove Theorem \ref{thm:gaussian_tails} we need  to rigorously establish the upper and lower bound.  We first investigate the upper bound.  As stated in Theorem \ref{thm:gaussian_tails} we will assume that $\Pr=1$ to simplify the calculations below.  We also assume that the forced modes $k$ are all even although this restriction could be removed with some additional calculations.  Moreover, we suppose that the horizontal domain length $L$ is an even integer. The basic approach to establish the upper bound on $\lambda_{\Pr}(\Ra,\zeta(\gamma))$ is to construct an admissible test function and to prove the lower bound we use general estimates.

\begin{lemma}\label{lem:gauss_tail}
For evenly forced modes $k$, and horizontal domain size $L$ an even integer, we find that
\begin{equation}
\lambda_{\Pr=1}(\Ra,\zeta(\gamma)) \leq \frac{1}{2}\min_k \left(\pi^2k^2 - \frac{\sqrt{2}}{4}\Ra^{1/2}\left[\sqrt{2}\pi k|\gamma_k| + 2\right]\right).
\end{equation}
\end{lemma}

\begin{proof}
For $m \in \mathbb{N} \setminus \{0\}$ (specified below), let $e_n = \sin(\pi n z)$, $g_m = \sin(2\pi m x/L)$ and $\overline{g}_m = \cos(2\pi m x/L)$.  Then
\begin{equation}
\f = (-g_m(x) e'_n(z),0,g_m'(x)e_n(z))^T
\end{equation}
is a divergence free vector field.  Then our trial flow field is given by $\u = Y\sqrt{\Ra}\f$ and the trial temperature fluctuation by $\theta = X\overline{g}_m(x)e_l(z)$.  
To belong to the set $\mathcal{M}$ (see \eqref{eq:variational_constraint} for the definition),
$(\u, \theta)$
 must satisfy $\|\theta\|_2^2 + \frac{1}{\Ra}\|\u\|_2^2 = 1$ (recalling the assumption $\Pr=1$) which becomes the following constraint on $X$ and $Y$:
\begin{equation*}
\frac{1}{4}X^2 + \pi^2\left(\frac{m^2}{L^2} + \frac{n^2}{4}\right) Y^2 = \frac{1}{L}.
\end{equation*}

Inserting these fields into the variational form of the definition of $\lambda$ we also arrive at
\begin{align*}
&\lambda_{\Pr=1}(\Ra,\zeta(\gamma)) \leq \frac{L}{4}\left[ X^2\left(\frac{4\pi^2m^2}{L^2} + \pi^2 l^2\right) + Y^2 \left(\frac{4\pi^2m^2}{L^2} + \pi^2n^2\right)^2\right]\\
&+ \Ra^{1/2}XY 2\pi m\int_0^1 \int_0^L \left(\zeta(\gamma) - 2\right) \sin(\pi n z) \sin(\pi l z) \cos^2(2\pi mx/L) dxdz.
\end{align*}
Recalling the definition \eqref{eq:define_zeta2} of $\zeta(\gamma)$ and the orthogonality of  trigonometric functions, we find that 
\begin{multline*}
\int_0^1 (\zeta(\gamma)-2)\sin(\pi nz) \sin(\pi l z)dz \\= \sum_{k \in \mathcal{N}}\frac{\sqrt{2}k\pi}{4}\gamma_k\left(\delta_{k-n-l} - \delta_{k-n+l}\right) - \frac{1}{2}\left(\delta_{n-l} - \delta_{n+l}\right)\,.
\end{multline*}
Fix any 
 $k\in\mathcal{N}$ and since, by assumption, $k$ is even (this choice is only to simplify the algebra at this point), depending on the sign of $\gamma_k$ we set $n= l = k/2$ or $n=-l=k/2$ such that the non-zero terms on the right-hand side have the same sign. Note that only the terms with the fixed $k$ can be non-zero.
   Moreover, because the cross term is odd in $Y$ we can always choose $Y$ so that this term is negative, and hence
\begin{align*}
\lambda_{\Pr=1}(\Ra,\zeta(\gamma)) &\leq \frac{L}{4}\left[X^2\left(\frac{4\pi^2m^2}{L^2}+ \frac{\pi^2k^2}{4}\right) + Y^2\left(\frac{4\pi^2m^2}{L^2} + \frac{\pi^2 k^2}{4}\right)^2\right]\\
&- \Ra^{1/2} XY \pi m \left(\frac{\sqrt{2}\pi k}{4}|\gamma_k| + \frac{1}{2}\right),
\end{align*}
where
\begin{equation*}
\frac{k}{2} = |l| = |m|,\quad \frac{1}{4}X^2 + \pi^2\left(\frac{m^2}{L^2}  + \frac{k^2}{16}\right)Y^2 = \frac{1}{L}.
\end{equation*}
Since $L$ and $k$ are even integers, 
we select $m= Lk/4$ 
and obtain an upper bound
\begin{equation}\label{eq:gaussian_tails_upper1}
\frac{4\lambda_{\Pr=1}(\Ra,\zeta(\gamma))}{L} \leq X^2\frac{\pi^2k^2}{2} + Y^2\frac{\pi^4k^4}{4} - \Ra^{1/2}XY\pi k \left(\frac{\sqrt{2}\pi k}{4}|\gamma_k| + \frac{1}{2}\right),
\end{equation}
with $X^2 + \frac{k^2\pi^2}{2} Y^2 = \frac{4}{L}$. After rescalling (note that $X$ and $Y$ are arbitrary) $X$ and $Y$ respectively by $\frac{2}{\sqrt{L}} X$ and  $\frac{2\sqrt{2}}{\pi k \sqrt{L}} Y$ we obtain $X^2 +  Y^2 = 1$
and
\begin{equation}\label{eq:gaussian_tails_upper2}
\lambda_{\Pr=1}(\Ra,\zeta(\gamma)) \leq X^2\frac{\pi^2k^2}{2} + Y^2\frac{\pi^2k^2}{2} - \sqrt{2\Ra} XY \left(\frac{\sqrt{2}\pi k}{4}|\gamma_k| + \frac{1}{2}\right),
\end{equation}
or equivalently, 
\begin{equation}\label{eq:gaussian_tails_upper3}
\lambda_{\Pr=1}(\Ra,\zeta(\gamma)) \leq  \frac{\pi^2k^2}{2} - \sqrt{2\Ra} XY \left(\frac{\sqrt{2}\pi k}{4}|\gamma_k| + \frac{1}{2}\right)\,.
\end{equation}
By standard arguments, the minimum of the right hand side is attained for $X = Y = \sqrt{2}/2$, and therefore 
\begin{equation}\label{eq:gaussian_tails_upper4}
\lambda_{\Pr=1}(\Ra,\zeta(\gamma)) \leq  \frac{\pi^2k^2}{2} - \frac{\sqrt{2\Ra}}{2} \left(\frac{\sqrt{2}\pi k}{4}|\gamma_k| + \frac{1}{2}\right)\,.
\end{equation}
for any $k \in \mathcal{N}$. After we take the minimum of the right-hand side in \eqref{eq:gaussian_tails_upper4} with respect to $k$, we obtain the desired upper bound. 
\end{proof}

The lower bound on the growth factor is much easier to obtain, but consequently also far less likely to be realized.  Using  Cauchy-Schwarz and Poincar\'e inequalities, definition of $\mathcal{M}$ implying 
$\|\theta\|_2^2 + \|\u\|_2^2 = 1$ we see that
\begin{align*}
&\lambda_{\Pr=1}(\Ra,\zeta(\gamma)) \\
&\geq \inf_{(\u,\theta)\in\mathcal{M}} \left\{\lambda_L(\|\theta\|_2^2 + \|\u\|_2^2) - \Ra^{1/2} \left(1+\sqrt{2}\pi\sum_{k\in\mathcal{N}}k|\gamma_k|\right)\|\u\|_2\|\theta\|_2\right\} 
\\
&\geq \lambda_L - \inf_{(\u,\theta)\in\mathcal{M}}
\Ra^{1/2} \left(1+\sqrt{2}\pi\sum_{k\in\mathcal{N}}k|\gamma_k|\right)
\frac{\|\u\|_2^2 +  \|\theta\|_2^2}{2} \\
&= 
\lambda_L - \frac{\Ra^{1/2}}{2} \left(1+\sqrt{2}\pi\sum_{k\in\mathcal{N}}k|\gamma_k|\right)\,,
\end{align*}
and the lower bound in Theorem \ref{thm:gaussian_tails} follows, 
where we used that $\lambda_L$ is the principal eigenvalue of the Laplace operator on the domain $\mathcal{D}$.

\section*{Acknowledgements}
We would like to thank G. Ahlers, C. R. Doering, B. Eckhardt, S. Friedlander, K. Lu, and J. Schumacher for helpful discussions and feedback.  In particular, C. R. Doering and B. Eckhardt both gave very positive and encouraging feedback on this work before their untimely passing, and this paper is partially dedicated to their memory.

This work was conceived and completed over numerous research visits.
In particular we are indebted to Mathematics departments at Brigham Young University, Universit\'e Libre de Bruxelles, Tulane University, the University of Virginia as well as the Mechanical and Aerospace Engineering department at Utah State University.   We would also like to thank  the Banff International Research Station which supported our work through a `Research in Peace fellowship' in October of 2016. JF was partially supported by the National Science Foundation under the grant DMS-1816408.  NEGH was partially supported by the National Science Foundation under the grants NSF-DMS-1313272, NSF-DMS-1816551, NSF-DMS-2108790. JPW was partially supported by the Simons Foundation travel grant under 586788 and the National Science Foundation under the grant NSF-DMS-2206762.

\bibliographystyle{elsarticle-num}%\bibliographystyle{elsarticle-harv}
% Note the spaces between the initials
\section*{References}
\bibliography{references.bib}

\end{document}